\documentclass[sigconf]{acmart}

\usepackage{booktabs} 
\usepackage{algorithm}
\usepackage{algorithmic}
\usepackage{threeparttable}
\usepackage{multirow}
\usepackage{newfloat}
\usepackage{listings}
\usepackage{xcolor}
\usepackage{subcaption}
\usepackage{balance}

\AtBeginDocument{%
  }

\DeclareCaptionStyle{ruled}{labelfont=normalfont,labelsep=colon,strut=off}
\floatstyle{ruled}
\newfloat{listing}{tb}{lst}{}
\floatname{listing}{Listing}

\graphicspath{{./figures}}

\newcommand{\ourm}{{ReCAP}}

\acmYear{2026}
\setcopyright{cc}
\setcctype{by}
\acmConference[KDD 2026] {Proceedings of the 32nd ACM SIGKDD Conference on Knowledge Discovery and Data Mining V.2}{August 9--13, 2026}{Jeju Island, Republic of Korea.}
\acmBooktitle{Proceedings of the 32nd ACM SIGKDD Conference on Knowledge Discovery and Data Mining V.2 (KDD 2026), August 9--13, 2026, Jeju Island, Republic of Korea}
\acmISBN{979-8-4007-2259-2/2026/08}
\acmDOI{10.1145/3770855.3817620}
\begin{document}

\title{Regime-Adaptive Continual Learning for Portfolio Management}

\author{Chaofan Pan}
\email{pan.chaofan@foxmail.com}
\orcid{0000-0001-5345-2746}
\affiliation{%
  \institution{Southwestern University of Finance and Economics}
  \city{Chengdu}
  \state{Sichuan}
  \country{China}
}

\author{Lingfei Ren}
\email{renlf@swufe.edu.cn}
\orcid{0000-0002-3756-3427}
\affiliation{%
  \institution{Southwestern University of Finance and Economics}
  \city{Chengdu}
  \state{Sichuan}
  \country{China}
}

\author{Linbo Xiong}
\email{xlb_beiqi@outlook.com}
\orcid{0009-0007-3402-9193}
\affiliation{%
  \institution{Southwestern University of Finance and Economics}
  \city{Chengdu}
  \state{Sichuan}
  \country{China}
}

\author{Yonghao Li}
\email{liyonghao@swufe.edu.cn}
\orcid{0000-0001-5653-4907}
\affiliation{%
  \institution{Southwestern University of Finance and Economics}
  \city{Chengdu}
  \state{Sichuan}
  \country{China}
}

\author{Wei Wei}
\email{weiwei@sxu.edu.cn}
\orcid{0000-0003-3963-2884}
\affiliation{%
  \institution{Shanxi University}
  \city{Taiyuan}
  \state{Shanxi}
  \country{China}
}

\author{Xin Yang}
\email{yangxin@swufe.edu.cn}
\orcid{0000-0002-0406-6774}
\authornote{Corresponding Author.}
\affiliation{%
  \institution{Southwestern University of Finance and Economics}
  \city{Chengdu}
  \state{Sichuan}
  \country{China}
}

\renewcommand{\shortauthors}{Chaofan Pan et al.}

\begin{abstract}
    Financial markets are inherently non-stationary, exhibiting frequent regime shifts and structural changes that render traditional Portfolio Management (PM) approaches ineffective.
    Existing remedies, such as rolling-window retraining and naive online fine-tuning, are hindered by high computational costs and insufficient knowledge utilization, respectively, resulting in low returns and limited adaptability.
    Continual learning (CL) offers a promising paradigm by enabling trading agents to accumulate and transfer knowledge across sequential tasks.
    In this paper, we propose \textbf{Re}gime-aware \textbf{C}ontinual \textbf{A}daptive \textbf{P}ortfolio management (\textbf{\ourm{}}), a novel framework that integrates CL into PM to address the challenges of dynamic financial environments.
    \ourm{} employs an adaptive regime detection module to segment historical market data into variable-length regimes, enabling regime-specific learning of policy vectors and the construction of a policy library.
    During continual trading, a regime-gate module adaptively combines policy vectors from the library based on the current market state, facilitating rapid adaptation to newly detected regimes.
    Only the regime-gate and the current regime's policy vector are continually updated to preserve useful knowledge effectively.
    Extensive experiments on five real-world datasets demonstrate that \ourm{} consistently outperforms popular baselines, achieving superior returns in long-term investment horizons and rapid adaptation to regime shifts.

\end{abstract}




\maketitle

\section{Introduction}
\label{sec: introduction}

The stock market, with a total market capitalization exceeding \$90 trillion, has attracted the attention of investors worldwide \cite{zhang2024reinforcement}. 
Portfolio Management (PM), which dynamically allocates capital proportions among different assets, plays a pivotal role in maximizing returns \cite{surtee2023novel,abate2024integration} and controlling risk for investors \cite{hoque2023time,gunjan2023brief}. 
With the recent advancements in deep learning and Reinforcement Learning (RL) \cite{oyewole2024predicting,shakya2023reinforcement}, quantitative trading has revolutionized PM by leveraging computational models to execute trades with unprecedented speed and precision \cite{agarwal2006algorithms,li2012online}.
However, financial markets are characterized by persistent non-stationarity and frequent structural shifts \cite{dhingra2024stock,otabek2024multi}, posing fundamental challenges for the design of robust and adaptive trading policies. 
While RL-based methods enable the trade agents to learn flexible trading policies through direct interaction with the market environment \cite{ye2020reinforcement,wang2021deeptrader}, these agents often encounter rapid performance degradation when deployed in real-world environments \cite{Katsikas2025Plasticity}. 
This phenomenon, commonly referred to as ``alpha decay'', arises as the market adapts to and arbitrages away profitable policies \cite{penasse2022understanding}, in line with the predictions of the Efficient Market Hypothesis \cite{sharpe1970efficient}. 
As a result, policies trained on offline data, under the assumption of stationarity, frequently fail to sustain their edge in the face of evolving market regimes and volatility shifts.
\begin{figure}
    \centering
    \includegraphics[width=0.47\textwidth]{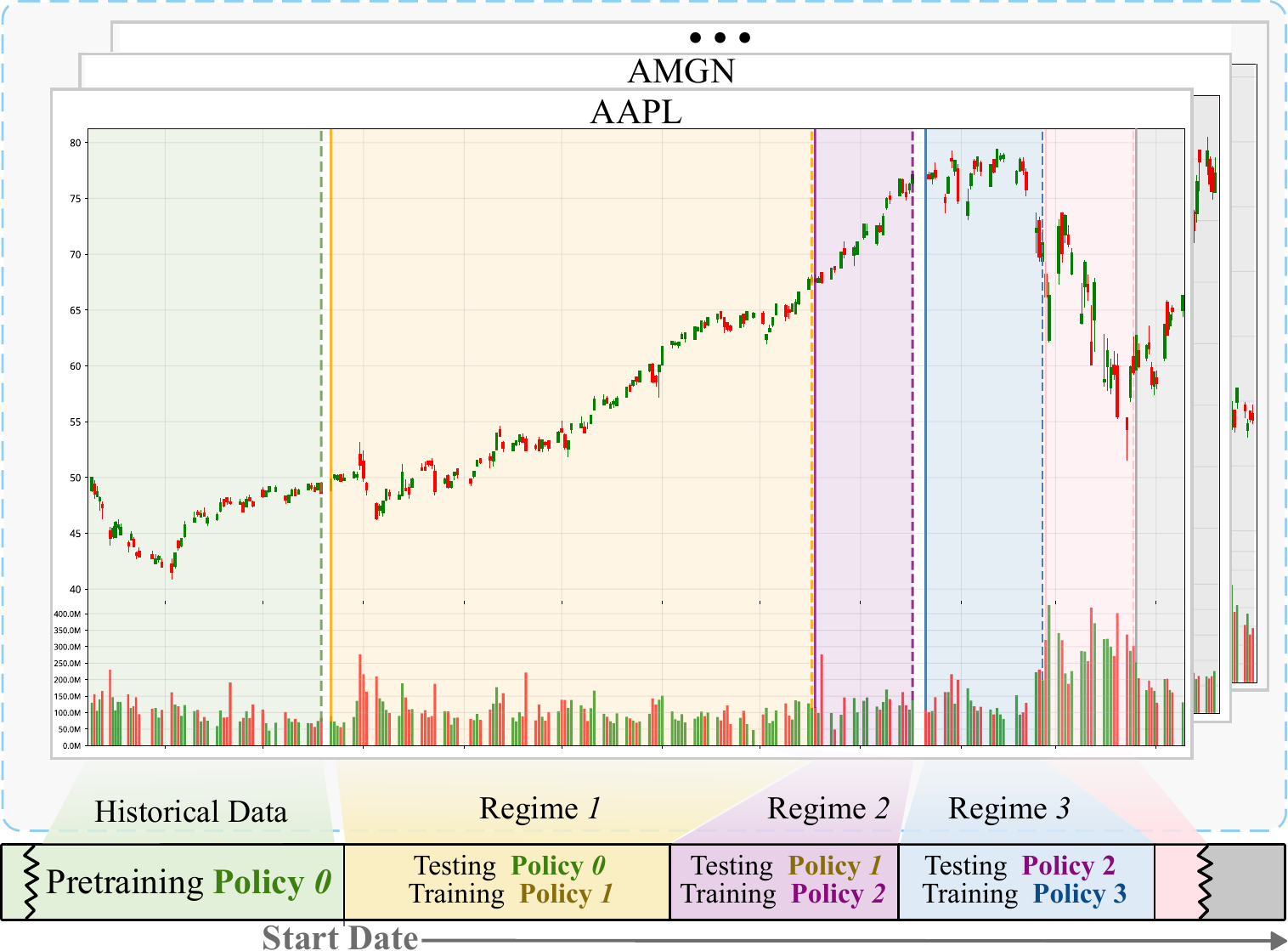}
    \caption{Illustration of the continual portfolio management problem.
    The multi-asset market evolves over time and is segmented into a sequence of regimes, each marked by distinct statistical properties. 
    Historical data is used for pre-training, while subsequent regimes are detected and used for continual training and testing in an alternating fashion.
    }
    \label{fig: 1}
\end{figure}

Continual Learning (CL) extends the static learning framework by enabling systems to learn across a sequence of tasks, incrementally accumulating and transferring knowledge over time \cite{khetarpal2022towards,abel2023definition}. 
CL has shown significant promise in various non-stationary domains, as it offers mechanisms to mitigate catastrophic forgetting and facilitate rapid adaptation to new tasks \cite{hafez2021behavior,geishauser2022dynamic,pan2025survey}. 
However, the application of CL in PM remains largely underexplored, and existing methods seldom exploit the unique characteristics of financial markets, such as regime-driven dynamics and the recurrence of market states \cite{schaller1997regime,ang2012regime,he2020global}. 
This gap motivates the development of new approaches that seamlessly integrate regime-aware CL with PM, empowering trading agents to achieve robust performance in real-world, ever-changing financial environments.

Building on the advances of CL, we propose to tackle PM as a CL problem under non-stationary market conditions. 
As illustrated in Figure \ref{fig: 1}, the problem of ``\textbf{continual portfolio management}'' is a sequence of non-stationary decision-making tasks.
The agent interacts with a multi-asset market environment whose statistical properties and underlying dynamics may shift abruptly over time due to structural changes, volatility spikes, or macroeconomic events. 
The entire market history is first used for offline pre-training, providing a foundation for the agent's initial knowledge. 
As the market evolves, the data is segmented into a series of regimes, each corresponding to a distinct period with relatively stable characteristics. 
At each regime boundary, the agent detects the shift, uses data from the previous regime for training, and then tests or deploys the learned policy in the subsequent regime.
The core objective is to maximize long-term returns by continuously adapting to new regimes, efficiently transferring knowledge from previously encountered market conditions.  
This CL process is more realistic and poses unique challenges: 
(1) How to detect and segment regime shifts in real time; 
(2) How to retain and reuse regime-specific expertise without incurring excessive computational or memory costs; 
and (3) How to dynamically combine historical and newly acquired knowledge to enable rapid adaptation to novel market conditions.

To address these challenges, we propose a novel framework named \textbf{\ourm{}} (\textbf{Re}gime-aware \textbf{C}ontinual \textbf{A}daptive \textbf{P}ortfolio management), which leverages advances in CL to enable robust and adaptive PM in non-stationary markets.
\ourm{} is designed to systematically detect market regime shifts, efficiently retain and reuse regime-specific expertise, and dynamically integrate historical and novel knowledge for rapid adaptation.
Firstly, our framework employs an adaptive regime detection module, which segments the historical market data into variable-length regimes. 
Secondly, the agent incrementally fine-tunes its policy for each regime to obtain a policy vector, which is then adaptively stored in a policy library. 
Thirdly, a regime-gate module is introduced during continual trading to assign attention weights to policy vectors in the library based on the current market state, enabling the agent to synthesize an effective trading policy tailored to the prevailing regime.
Only the regime-gate and the current regime's policy vector are updated continually, which supports efficient knowledge accumulation.
By introducing an adaptive regime detection module, policy vectors, a policy library, and a regime-gate module, \ourm{} provides a unified solution to the challenges of continual PM in real-world, dynamic financial environments.

In summary, this paper makes the following contributions:
\begin{itemize}
    \item We formulate and address the continual PM problem, explicitly modeling the non-stationary nature of real-world financial markets as a sequence of regime-aware decision-making tasks, and highlighting the necessity of continual adaptation for robust trading performance.
    \item We propose \ourm{}, a novel continual PM framework that integrates adaptive regime detection, regime-specific policy learning and maintenance, and regime-aware policy composition into a unified framework for efficient knowledge accumulation and rapid adaptation.
    \item We conduct extensive empirical studies on five real-world datasets, covering US and Japanese stock markets as well as commodity ETFs, demonstrating that \ourm{} consistently outperforms strong baselines across multiple metrics.
\end{itemize}

\section{Related Works}
\label{sec: related_work}
Portfolio Management is a critical task in finance, involving the allocation of capital across multiple assets to achieve specific investment objectives such as maximizing returns or minimizing risk.
Pioneering studies, including Universal Portfolios (UP) \cite{blum1997universal}, Constant Rebalanced Portfolios (CRP) \cite{helmbold1998line}, Exponential Gradient (EG) \cite{helmbold1998line}, M0 \cite{borodin2000competitive}, Anti-Correlation (Anticor) \cite{borodin2003can}, and Online Newton Step (ONS) \cite{agarwal2006algorithms}, have been proposed to achieve a high-profit investment.
These methods typically rely on classical financial theories such as mean-variance optimization \cite{erlich2010mean} and factor models \cite{giglio2022factor}, which assume relatively stable market dynamics and often require strong assumptions about asset return distributions.
While these methods provide foundational insights, their performance can deteriorate in complex, non-stationary financial environments \cite{ye2020reinforcement,zheng2024cross}.
Moreover, traditional methods typically lack the flexibility to incorporate market impact and evolving market information in a unified framework.

RL has emerged as a powerful alternative paradigm for PM by formulating the investment decision process as a sequential decision-making problem \cite{shakya2023reinforcement}. 
Unlike supervised learning methods \cite{bai2018empirical}, RL agents learn adaptive trading policies through direct interaction with the market environment, optimizing long-term, risk-adjusted returns while naturally incorporating transaction costs and constraints. 
Early works, such as EIIE \cite{jiang2017deep} and FinRL \cite{liu2021finrl}, laid the groundwork for applying deep RL algorithms to PM. 
More recent developments have focused on enhancing expressiveness and robustness. 
SARL \cite{ye2020reinforcement} introduced state augmentation techniques to better capture asset dynamics, while HRPM \cite{wang2021commission} proposed hierarchical RL frameworks to separate long-term PM from short-term execution, effectively reducing trading costs. 
Advances in graph neural networks \cite{Farzan2021DeepPocket} and multi-agent systems \cite{lee2020maps} further improved the modeling of asset correlations.

Despite these successes, existing RL-based PM methods predominantly assume stationary or slowly varying market conditions and generally rely on offline training or periodic retraining with fixed windows. 
It is essential to integrate CL principles into the RL-based method, enabling agents to adapt to dynamic financial markets.
Although two recent studies have attempted to apply continual RL to algorithmic trading, they primarily focus on short-term, single-asset environments with fixed intervals and lack sufficient comparison \cite{Katsikas2024BiDirectional,Katsikas2025Plasticity}.
Additionally, CL methods have not been explored in the context of PM.
To address these gaps, we propose a novel CL framework for PM on long-term, multi-asset environments with adaptive regime detection, enabling rapid adaptation to non-stationary market conditions.

\section{Problem Formulation}
\label{sec: preliminaries}

\paragraph{Definitions.}
Let $N$ denote the number of tradable assets in the market. 
At each time step $t$, the market observation is denoted as $\mathbf{X}_t = \{x_{t,1}, \ldots, x_{t,N}\}$, where $x_{t,i}$ represents the feature vector (e.g., OHLCV and technical indicators) for asset $i$. 
The portfolio allocation at time $t$ is represented by the weight vector $\mathbf{w}_t = [w_{t,0}, w_{t,1},\\ \ldots, w_{t,N}]$, where $w_{t,i}$ is the proportion of capital allocated to asset $i$. 
The portfolio value at time $t$ is denoted as $V_t$.

\paragraph{Portfolio Management.}
The trading agent interacting with a financial market environment can be modeled as a Markov Decision Process (MDP) $(\mathcal{S}, \mathcal{A}, \mathcal{R}, \mathcal{T}, \gamma)$ \cite{ijcai2020p641,liu2021finrl,zhang2022cost}, where $\mathcal{S}$ is the state space, $\mathcal{A}$ is the action space, $\mathcal{R}: \mathcal{S} \times \mathcal{A} \to \mathbb{R}$ is the reward function, $\mathcal{T}: \mathcal{S} \times \mathcal{A} \times \mathcal{S} \to [0,1]$ is the transition dynamics, and $\gamma\in [0,1)$ is the discount factor.
At each discrete time step $t$, the trading agent observes the trading state $\mathbf{s}_t \in \mathcal{S}$, follows a policy $\pi: \mathcal{S} \to \mathcal{A}$ to select a portfolio allocation action $\mathbf{a}_t \equiv \mathbf{w}_t \in \mathcal{A}$, and receives a reward $r_t = \mathcal{R}(\mathbf{s}_t, \mathbf{a}_t)$ reflecting the portfolio return.
The state transitions to the next state $\mathbf{s}_{t+1}$ according to the transition dynamics.

Specifically, the state $\mathbf{s}_t = f(X_t \cup \mathcal{H}_t)$ consists of the latest multi-asset market features, where $f$ and $\mathcal{H}_t$ denote a feature extraction function and optional historical features, respectively.
The action $\mathbf{w}_t$ is the portfolio weight vector at time $t$, which is subject to the constraints $\sum_{i=0}^{N} w_{t,i} = 1$ and $w_{t,i} \geq 0$.
The immediate reward is the log portfolio return $r_t = \log \left( \frac{V_t}{V_{t-1}} \right)$, where $V_t$ is the portfolio value at time $t$.
The transition dynamics are implicitly defined by the market evolution, which is influenced by the agent's actions and external factors.

\paragraph{Continual Portfolio Management.}
We formulate the continual PM problem as an MDP under regime-shifting market conditions.
In the real world, the financial market is inherently non-stationary and can be viewed as a sequence of regimes, $\{\tau_1, \tau_2, \ldots, \tau_M\}$, where each regime $\tau_k$ corresponds to a contiguous time segment with relatively stable statistical properties.
As time progresses, regime boundaries and their characteristics are typically unknown in advance.
To model this, we formulate the continual PM as a CRL problem with dynamic MDP: $(\mathcal{S}, \mathcal{A}, \mathcal{R}_k, \mathcal{T}_k, \gamma)$.
Within each regime $\tau_k$, the underlying transition dynamics $\mathcal{T}_k$ and reward function $\mathcal{R}_k$ may differ, reflecting changes in market volatility or return distributions. 
Thus, the agent faces a sequence of related PM tasks, where the optimal policy may shift as the regime changes.

The CRL process alternates between adaptation and deployment.
After a regime boundary is detected, the agent trains on the completed regime $\tau_{k-1}$ under $\mathcal{T}_{k-1}$ and $\mathcal{R}_{k-1}$ to capture its dynamics.
The adapted policy $\pi_{k-1}$ is then deployed to make portfolio allocation decisions in the subsequent regime $\tau_k$ until the next regime shift is detected.
The objective is to maximize cumulative portfolio return over the entire time horizon:
\begin{equation}
\max_{\{\pi_{k-1}\}_{k=1}^{M}} \mathbb{E} \left[ \sum_{k=1}^M \sum_{t=0}^{|\tau_k|-1} r^{(k)}_t \right],
\end{equation}
where $\pi_{k-1}$ denotes the policy used in $\tau_k$, $|\tau_k|$ denotes the number of trading steps in regime $\tau_k$, and $r^{(k)}_t = \mathcal{R}_{k}(\mathbf{s}_t, \mathbf{a}_t)$ is the reward at time $t$ under regime-specific dynamics $\mathcal{T}_k(\cdot)$.

\section{Method}
\label{sec: method}
This section presents our proposed framework for continual PM.
We first provide an overview of \ourm{}, followed by detailed descriptions of its core components: the \textit{Adaptive Regime Detection module} (ARD), the \textit{policy vector and policy library}, and the \textit{Regime-Gate Module} (RGM).
The overall workflow of \ourm{} is summarized in Algorithm \ref{alg:pretrain} and Algorithm \ref{alg:continual}.

\begin{figure*}[htbp]
    \centering
    \includegraphics[width=0.75\linewidth]{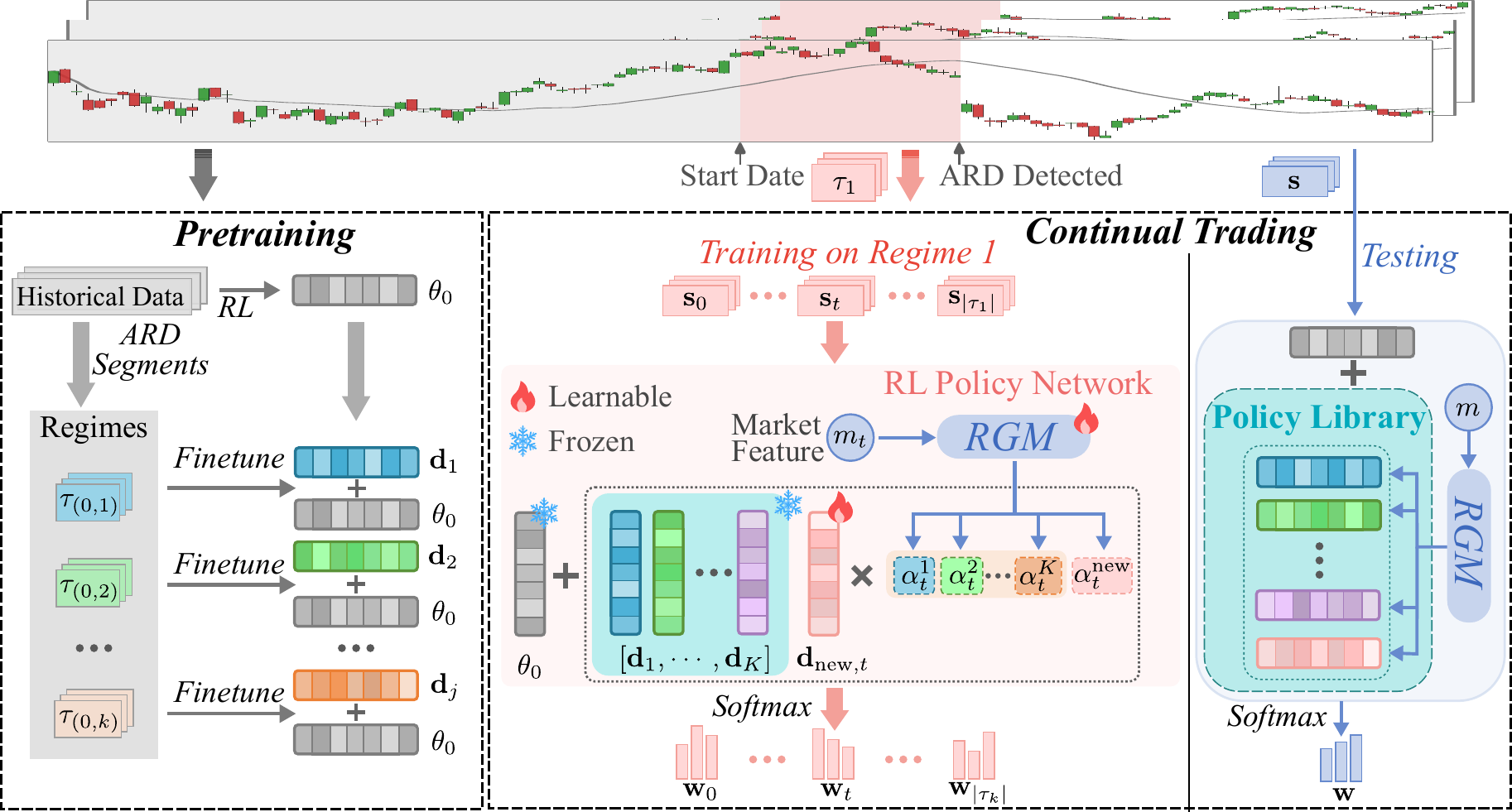}
    \caption{Overview of the \ourm{}'s workflow. 
The offline pretraining stage produces a base policy. 
The Adaptive Regime Detection module (ARD) segments historical data into distinct regimes. 
Each regime is used to fine-tune the base policy, yielding regime-specific policy vectors and building the policy library.
In the continual trading stage, real-time ARD triggers the Regime-Gate Module (RGM) to dynamically combine stored policy vectors with the current policy vector, producing the current policy network. 
A new policy vector is trained and adaptively merged into the policy library.
During testing, a daily state is processed by a dynamic policy network consisting of the base policy, policy library, and RGM to obtain a portfolio weight.
}
    \label{fig: framework}
\end{figure*}

\subsection{Overview}
Figure \ref{fig: framework} illustrates the overall workflow of \ourm{}.
It consists of two main stages: \textit{pretraining} and \textit{continual trading}.

\paragraph{Pretraining Stage.}  
We begin by training a base PM policy parameterized by $\theta_0$ on extensive historical market data covering a long time horizon. 
This offline pretraining employs a standard RL algorithm to learn a robust initial policy that captures general market dynamics.
Next, the ARD analyzes the historical data to identify structural change points and partitions the data into multiple regimes $\{\tau_{(0,1)},\cdots, \tau_{(0,K)}\}$, each representing a period with relatively stable market characteristics.
For each detected regime $\tau_{(0,j)}$, the base policy $\theta_0$ is fine-tuned to adapt to regime-specific dynamics, producing a policy $\theta_j$ and a policy vector $\mathbf{d}_j = \theta_j - \theta_0$.
Collectively, these vectors are merged to form a policy library $\mathbf{D} = [\mathbf{d}_1, \cdots, \mathbf{d}_K]$, which is stored for reuse.

\paragraph{Continual Trading Stage.}  
During continual trading, the framework continuously monitors the market and applies the ARD in real time to detect regime shifts.
If a regime boundary is detected, the completed regime segment is used for adaptation to produce a new policy vector $\mathbf{d}_{\text{new}}$, which is then available for subsequent trading.
Specifically, at each time step $t$, the policy network receives the asset-level trading state $\mathbf{s}_t$, while the RGM receives a market-level regime feature vector $\mathbf{m}_t$ constructed from market indicators and aggregated asset information.
The RGM outputs a weight vector $\boldsymbol{\alpha}_t$ over the policy library and, during adaptation, the current regime vector.
The effective policy parameter at time $t$ is then composed as:
\begin{equation}
    \label{eq:policy_composition}
    \theta_t = \theta_0 + \tilde{\mathbf{D}} \boldsymbol{\alpha}_t, \quad \tilde{\mathbf{D}} = [\mathbf{D}, \mathbf{d}_{\text{new},t}],
\end{equation}
where $\mathbf{d}_{\text{new},t}$ is the policy vector being learned for the current regime at time $t$, and $\mathbf{d}_{\text{new},0}$ is initialized as zero.
The portfolio weight $\mathbf{w}_t$ is computed as:
\begin{equation}
    \label{eq:portfolio_weights}
    \begin{aligned}
    \mathbf{w}_t & = \frac{\exp(\mathbf{o}_t)}{\sum_{i=0}^N \exp(o_{t,i})}, \quad \mathbf{o}_t = f_{\theta_t}(\mathbf{s}_t),
    \end{aligned}
\end{equation}
where $\mathbf{s}_t$ is the current state, $f_{\theta_t}$ is the policy network parameterized by $\theta_t$.
Only the RGM parameters $\phi$ and $\mathbf{d}_{\text{new}}$ are updated during training, while $\theta_0$ and $\mathbf{D}$ remain fixed to preserve the learned knowledge.
After training, the new policy vector $\mathbf{d}_{\text{new}}$ is adaptively merged into the policy library $\mathbf{D}$, accumulating regime-specific knowledge over time.

In testing, the ARD and the RGM operate jointly with the base policy $\theta_0$ and the policy vector library $\mathbf{D}$ to generate dynamic policy parameters for each state.
The portfolio weight vector $\mathbf{w}_t$ is then computed from the current trading state $\mathbf{s}_t$ using the composed policy.
This design enables efficient zero-shot inference and robust adaptation to evolving market regimes.

\subsection{Adaptive Regime Detection Module (ARD)}
\label{subsec: adaptive regime detection module}
A key challenge in continual PM is to identify and segment the market into regimes.
Accurate detection enables the agent to localize adaptation and facilitates the building of regime-specific expertise.
To achieve this, we propose the ARD to segment the market into regimes with approximately stationary dynamics.

Given a sequence of market observations $\{\mathbf{X}_t\}_{t=1}^L$ with length $L$, the goal is to partition the market into $M$ regimes $\{\tau_1, \tau_2, \ldots, \tau_M\}$ such that each regime $\tau_k = [t_k^{\mathrm{start}}, t_k^{\mathrm{end}}]$ corresponds to a time interval with approximately stationary dynamics.
Let $\mathbf{m}_t$ denote the market-level feature vector at time $t$, which includes indicators such as VIX, turbulence, and aggregated asset statistics.
The regime detection task reduces to identifying a set of change points $\mathcal{C} = \{c_1, c_2, \ldots, c_{M-1}\}$ such that the distribution of $\mathbf{m}_t$ changes significantly at each $c_i$.

We employ a variant of the cumulative sum algorithm \cite{warner2022cusum} to detect abrupt changes in the distribution of selected market features. 
For a chosen regime-sensitive scalar feature $u_t$, the statistic is defined as:
\begin{equation}
    \label{eq:cusum_statistic}
    S_t = \max(0, S_{t-1} + (u_t - \mu_0) - \kappa),
\end{equation}
where $\mu_0$ is a reference mean estimated from a historical reference window, and $\kappa > 0$ is a drift parameter controlling detection sensitivity.
A regime change is signaled whenever $S_t$ exceeds a predetermined threshold $h$.
After each detected change point, the statistic is reset to zero and $\mu_0$ is re-estimated on the subsequent reference window.
This procedure is applied in parallel to multiple features, and the union of detected change points is used for segmentation:
\begin{equation}
    \label{eq:change_points_union}
    \mathcal{C} = \bigcup_{u \in \mathcal{U}} \mathcal{C}_u,
\end{equation}
where $\mathcal{U}$ is the set of market-level features in $\mathbf{m}_t$, and $\mathcal{C}_u$ denotes the set of change points detected on feature $u$.

Given the set of change points $\mathcal{C}$, the market data is partitioned into regimes:
\begin{equation}
    \label{eq:regime_partitioning}
    \tau_1 = [1, c_1], \cdots, \tau_M = [c_{M-1}+1, L].
\end{equation}
Each regime $\tau_k$ is then treated as a distinct task for regime-specific policy adaptation.
During the continual trading stage, the ARD operates in an online fashion, sequentially updating the statistic $S_t$ and flagging a new regime whenever a change point is detected.

\subsection{Policy Vector and Policy Library}
\label{subsec: policy vector library}
In CL, leveraging knowledge from previous tasks in dynamic environments is crucial to enhance agent performance in subsequent tasks.
Therefore, we introduce the concept of policy vector and build a policy library to facilitate efficient knowledge retention and flexible reuse across different market regimes, inspired by model editing techniques \cite{ilharco2023editing}.
Then, we store regime-specific knowledge as a set of policy vectors and adaptively maintain the library through merging and pruning.

For each detected regime $\tau_{(0,j)}$, we initialize the policy parameters as $\theta_0$ and fine-tune them on regime-specific data to obtain adapted parameters $\theta_j$. 
The policy vector is then defined as the learned parameters that adapt the base policy to a specific regime:
\begin{equation}
    \label{eq:policy_vector}
    \mathbf{d}_j = \theta_j - \theta_0.
\end{equation}
The set of all such vectors forms the policy library
$\mathbf{D} = [\mathbf{d}_1, \ldots, \mathbf{d}_K] \in \mathbb{R}^{P \times K}$,
where $P$ is the dimension of the policy parameter space.
A linear combination of multiple policy vectors enables the model to reuse policy from previous regimes \cite{kim2025testtime}. 
This highlights the potential of policy vectors to facilitate continual adaptation.

To maintain a compact and informative policy library, we employ an adaptive merging mechanism that dynamically updates the library based on the similarity of policy vectors.
After generating all policy vectors from pretrain data, we compute the similarity between every pair of vectors. 
If the similarity between two vectors $\mathbf{d}_i$ and $\mathbf{d}_j$ exceeds a threshold $\delta_{\text{s}}$, they are merged by averaging:
\begin{equation}
    \label{eq:vector_merging}
    \mathbf{d}_{\text{m}} = \frac{1}{2}(\mathbf{d}_i + \mathbf{d}_j).
\end{equation}
Moreover, non-informative policy vectors whose $\ell_2$-norm is very small are discarded, as they do not contribute significantly to the policy adaptation process.
When a new regime is encountered and a new policy vector $\mathbf{d}_{\text{new}}$ is learned, it is first temporarily added to $\mathbf{D}$.
If the average attention weight from the RGM assigned to $\mathbf{d}_{\text{new}}$ is close to zero, or if its $\ell_2$-norm is very small, the new vector is discarded.
This indicates that the new regime does not require significant adaptation from the base policy and the policy library, and a combination of previous policy vectors is sufficient.
Otherwise, $\mathbf{d}_{\text{new}}$ is merged with existing vectors in $\mathbf{D}$ using the same similarity-based criterion as above.
This mechanism reduces redundancy and ensures that the policy library retains only distinct and useful regime adaptations.

\subsection{Regime-Gate Module (RGM)}
\label{subsec: regime-gate module}
The RGM is designed to dynamically synthesize an effective trading policy by adaptively weighting the policy vectors in the library according to the current market regime.
At each time step $t$, the market-level feature vector $\mathbf{m}_t$ is fed into a neural network parameterized by $\phi$:
\begin{equation}
    \label{eq:gate_mlp}
    \boldsymbol{\alpha}_t = \mathrm{Softmax}(g_\phi(\mathbf{m}_t)),
\end{equation}
where $\boldsymbol{\alpha}_t = [\alpha^1_t, \alpha^2_t, \ldots, \alpha^K_t]$ is a weight vector, $K$ is the number of policy vectors in the library, and $\boldsymbol{\alpha}_t$ is the attention weight vector over the policy library.
During inference, the current policy parameter is composed as:
\begin{equation}
    \label{eq:policy_composition_final}
    \theta_t = \theta_0 + \mathbf{D} \boldsymbol{\alpha}_t.
\end{equation}

During training on each regime, the output head of the RGM is first expanded to output a new weight $\alpha^\text{new}_t$ for the current regime policy vector $\mathbf{d}_{\text{new}, t}$.
Only the regime-gate parameters $\phi$ and the current regime's policy vector $\mathbf{d}_{\rm{new}}$ are updated via RL algorithms, while $\theta_0$ and $\mathbf{D}$ remain fixed.
After training, the new policy vector $\mathbf{d}_{\text{new}}$ is adaptively merged into the policy library $\mathbf{D}$, as described in Section \ref{subsec: policy vector library}.
If the new policy vector is discarded, the output head of the RGM is pruned to output only the weights for the existing policy vectors in the library.

\begin{algorithm}[t]
    \caption{Pretraining for \ourm{}}
    \label{alg:pretrain}
    \textbf{Input}: Historical market data $\{\mathbf{X}_t\}_{t=1}^T$ \\
    \textbf{Output}: Base policy parameter $\theta_0$, policy library $\mathbf{D}$ \\
    \begin{algorithmic}[1]
    \STATE \textit{// Step 1: Offline base policy learning}
    \STATE Train policy network $f_{\theta_0}$ on $\{\mathbf{X}_t\}_{t=1}^T$ using standard RL algorithm;
    \STATE \textit{// Step 2: Regime segmentation}
    \STATE Apply ARD to $\{\mathbf{X}_t\}_{t=1}^T$ to detect change points $\mathcal{C}$ and partition data into regimes $\{\tau_{(0,1)}, \ldots, \tau_{(0,K)}\}$;
    \STATE Initialize policy library $\mathbf{D} \leftarrow \emptyset$;
    \FOR{each regime $\tau_{(0,j)}$}
        \STATE Initialize $\theta_j \leftarrow \theta_0$;
        \STATE Fine-tune $\theta_j$ on data in $\tau_{(0,j)}$ via RL algorithm;
        \STATE Compute policy vector $\mathbf{d}_j$ by Eq. (\ref{eq:policy_vector});
        \STATE Add $\mathbf{d}_j$ to $\mathbf{D}$;
    \ENDFOR
    
    \STATE \textit{// Step 3: Policy vector merging}
    \STATE For all pairs $(\mathbf{d}_i, \mathbf{d}_j)$ in $\mathbf{D}$, merge by Eq. (\ref{eq:vector_merging}) if similarity $> \delta_{\text{s}}$;
    \STATE Remove policy vectors with small $\ell_2$-norm from $\mathbf{D}$;
    \STATE \textbf{return} $\theta_0$, $\mathbf{D}$;
    \end{algorithmic}
\end{algorithm}

\begin{algorithm}[t]
    \caption{Continual Trading with \ourm{}}
    \label{alg:continual}
    \textbf{Input}: Daily market data $\mathbf{X}$, base policy $\theta_0$, policy library $\mathbf{D}$, RGM parameters $\phi$, data buffer $\mathcal{B}$, ARD statistics $S$\\
    \textbf{Output}: Portfolio weights $\mathbf{w}$, updated $\mathbf{D}$, $\phi$, $\mathcal{B}$ and $S$\\
    \begin{algorithmic}[1]
        \STATE \textit{// Step 1: Portfolio weights computation}
        \STATE Obtain trading state $\mathbf{s}_t$ and market-level features $\mathbf{m}_t$;
        \STATE Compute policy weights $\boldsymbol{\alpha}$ by Eq. (\ref{eq:gate_mlp});
        \STATE Compose policy parameters $\theta$ by Eq. (\ref{eq:policy_composition_final});
        \STATE Compute portfolio weights $\mathbf{w}$ by Eq. (\ref{eq:portfolio_weights});
        \STATE \textit{// Step 2: Regime detection}
        \STATE Update ARD statistics $S$ by Eq. (\ref{eq:cusum_statistic});
        \IF{regime change detected}
            \STATE \textit{// Step 3: Regime-specific training}
            \STATE Initialize $\mathbf{d}_{\text{new}} \leftarrow \mathbf{0}$;
            \STATE Expand the output head of RGM;
            \STATE Update $\mathbf{d}_{\text{new}}$ and $\phi$ by RL on $\mathcal{B}$, keeping $\theta_0$, $\mathbf{D}$ fixed;
            
            \STATE \textit{// Step 4: Policy vector merging and library update}
            \IF{average weight of $\mathbf{d}_{\text{new}}$ is small or $\|\mathbf{d}_{\text{new}}\|_2$ is small}
                \STATE Discard $\mathbf{d}_{\text{new}}$;
                \STATE Prune the output head of RGM;
            \ELSIF{the similarity between $\mathbf{d}_{\text{new}}$ and any policy vector in $\mathbf{D}$  $> \delta_{\text{s}}$}
                \STATE Merge $\mathbf{d}_{\text{new}}$ into $\mathbf{D}$ by Eq. (\ref{eq:vector_merging});
                \STATE Prune the output head of RGM;
            \ELSE
                \STATE $\mathbf{D} \leftarrow [\mathbf{D},\mathbf{d}_{\text{new}}]$;
            \ENDIF
            \STATE Reset $\mathbf{d}_{\text{new}} \leftarrow \mathbf{0}$, clear $\mathcal{B}$;
        
        \ELSE
            \STATE Store $(\mathbf{s}_t, \mathbf{m}_t, r_t)$ into buffer $\mathcal{B}$;
        \ENDIF
        \STATE \textbf{return} portfolio weights $\mathbf{w}$, updated $\mathbf{D}$, $\phi$, $\mathcal{B}$, and $S$;
    \end{algorithmic}
\end{algorithm}

\section{Experiments}
\label{sec: experiments}

\subsection{Experimental Setup}
\paragraph{Datasets.}

\begin{table}[t]
\centering
\renewcommand{\arraystretch}{0.7}
\caption{Information about five datasets.}
\label{tab:datasets}
\resizebox{0.48\textwidth}{!}{
\begin{tabular}{lccc}
\toprule
Dataset & Assets & Training Period & Evaluation Period \\
\midrule
DOW30 & 29 & 2008-05-01 to 2020-04-30 & 2020-05-01 to 2025-04-29 \\
NAS100 & 73 & 2008-05-01 to 2020-04-30 & 2020-05-01 to 2025-04-29 \\
SP500 & 398 & 2008-05-01 to 2020-04-30 & 2020-05-01 to 2025-04-29 \\
NIKKEI30 & 29 & 2008-05-01 to 2020-04-30 & 2020-05-01 to 2025-04-29 \\
COMMODITY\_ETF & 7 & 2008-05-01 to 2020-04-30 & 2020-05-01 to 2025-04-29 \\
\bottomrule
\end{tabular}
}
\end{table}

\begin{table*}
    \renewcommand{\arraystretch}{0.7}
    \setlength{\abovecaptionskip}{0.1cm}
    \centering
    \caption{Performance comparison of thirteen PM methods and \ourm{} on DOW30, NAS100 and SP500. 
    The mean and standard deviation of the results are reported for RL-based methods. 
    Results in bold show the best results on each dataset.}
    \begin{threeparttable}
    \footnotesize
    \setlength{\tabcolsep}{1mm}{
    \resizebox{0.99\linewidth}{!}{
    \begin{tabular}{cl|ccc|ccc|ccc} 
    \toprule
    \multicolumn{2}{c|}{Dataset} & \multicolumn{3}{c}{\textbf{DOW30}} & \multicolumn{3}{c}{
    \textbf{NAS100}} & \multicolumn{3}{c}{\textbf{SP500}}\\
    
    \midrule
    \multicolumn{2}{c|}{Method} & CR\%$\uparrow$ & SR$\uparrow$ & MDD\%$\downarrow$ & CR\%$\uparrow$ & SR$\uparrow$ & MDD\%$\downarrow$ & CR\%$\uparrow$ & SR$\uparrow$ & MDD\%$\downarrow$ \\
    \midrule
    
    \multirow{6}{*}{\begin{tabular}[c]{@{}c@{}}Rule-\\based\end{tabular}} 
    
    & 	B\&H	& 84.10  & 0.88 & 20.91 & 123.41  & 0.92 & 25.37 & 111.00  & 0.92 & 19.84 \\
    & 	CRP	& 62.41 & 0.74  & 23.01  & 83.10  & 0.73  & 30.76 & 74.44 & 0.74 & 24.51 \\
    & 	EG	& 62.37  & 0.74  & 23.03  & 82.92  & 0.73 & 30.94 & 74.28 & 0.74 & 24.55 \\
    & 	UP	& 62.33  & 0.74  & 23.00 & 83.16  & 0.73  & 30.79 & 74.37 & 0.74 & 24.51 \\
    & 	OLMAR	& -68.38  & -0.47  & 82.99 & -90.98  & -0.72 & 95.59 & -93.27 & -0.78  & 96.38 \\
    & 	WMAMR	& -5.57  & 0.12 & 60.13  & -89.98  & -0.87 & 94.27 & -61.50 & -0.20 & 78.76 \\
    \midrule
    
    \multirow{7}{*}{\begin{tabular}[c]{@{}c@{}}RL-\\based\end{tabular}} 
    
    & 	A2C	& 80.28 $\pm$ 11.37 & 0.85 $\pm$ 0.08 & 20.59 $\pm$ 2.13 & 126.10 $\pm$ 5.34  & 0.93 $\pm$ 0.02  & 24.95 $\pm$ 0.77  & 112.08 $\pm$ 3.15 & 0.92 $\pm$ 0.02 & 19.98 $\pm$ 0.57 \\
    & 	PPO	& 84.34 $\pm$ 1.49 & 0.88 $\pm$ 0.01 & 20.92 $\pm$ 0.18 & 123.54 $\pm$ 1.30  & 0.92 $\pm$ 0.01  & 25.37 $\pm$ 0.17 & 111.08 $\pm$ 0.22 & 0.92 $\pm$ 0.00 & 19.82 $\pm$ 0.01 \\
    & 	SAC & 85.86 $\pm$ 5.23 & 0.89 $\pm$ 0.04 & 20.84 $\pm$ 1.02 & 123.54 $\pm$ 7.44 & 0.92 $\pm$ 0.03  & 25.27 $\pm$ 0.94  & 111.55 $\pm$ 2.40 & 0.92 $\pm$ 0.01 & 19.98 $\pm$ 0.64 \\
    & 	EIIE &	83.17 $\pm$ 0.95 & 0.87 $\pm$ 0.01 & 20.92 $\pm$ 0.10 & 123.33 $\pm$ 0.32 & 0.92 $\pm$ 0.00 & 25.35 $\pm$ 0.04 & 110.99 $\pm$ 0.02  & 0.92 $\pm$ 0.00 & 19.84 $\pm$ 0.00\\
    & 	SARL	& 78.69 $\pm$ 1.85 & 0.84 $\pm$ 0.02 &  21.07 $\pm$ 0.29 & 111.87 $\pm$ 0.78  & 0.86 $\pm$ 0.00 & 25.37 $\pm$ 0.19 & 103.93 $\pm$ 0.29 & 0.89 $\pm$ 0.00 & 19.83 $\pm$ 0.07 \\
    & 	Cross-Insight & 90.31 $\pm$ 9.30 & 0.89 $\pm$ 0.06 & 20.85 $\pm$ 1.09 & 125.37 $\pm$ 4.93 & 0.91 $\pm$ 0.02 & 25.73 $\pm$ 0.96  & 110.83 $\pm$ 2.54 & 0.92 $\pm$ 0.01 & 19.92 $\pm$ 0.64 \\
    & AlphaGAT & 73.45$\pm$0.00 & 0.80$\pm$0.00 & 21.46$\pm$0.00 & $123.38 \pm 0.00$ & $0.91 \pm 0.00$ & $25.36 \pm 0.00$  & $111.02\pm0.01$ & $0.92\pm0.00$ & $19.84\pm0.00$ \\
    
    \cmidrule{2-11}
    & 	\textbf{\texttt{\ourm{}}}	& \textbf{96.76 $\pm$ 2.26} & \textbf{1.00 $\pm$ 0.02} & \textbf{17.58 $\pm$ 0.48}  & \textbf{164.89 $\pm$ 2.54} & \textbf{1.14 $\pm$ 0.01} & \textbf{23.86 $\pm$ 0.35} & \textbf{145.02 $\pm$ 1.40} & \textbf{1.14 $\pm$ 0.01} & \textbf{19.19 $\pm$ 0.13} \\
    
    \bottomrule
    
    \end{tabular}
    }}
    \end{threeparttable}
    \label{table:comparison}
    \end{table*}

\begin{table*}
    \renewcommand{\arraystretch}{0.7}
    \setlength{\abovecaptionskip}{0.1cm}
    \centering
    \caption{Performance comparison on NIKKEI30 and COMMODITY\_ETF.}
    \label{tab:comparison_new}
    \footnotesize
    \setlength{\tabcolsep}{1mm}{
    \resizebox{0.7\linewidth}{!}{
    \begin{tabular}{cl|ccc|ccc}
    \toprule
    \multicolumn{2}{c|}{Dataset} & \multicolumn{3}{c}{\textbf{NIKKEI30}} & \multicolumn{3}{c}{\textbf{COMMODITY\_ETF}} \\
    \midrule
    \multicolumn{2}{c|}{Method} & CR\%$\uparrow$ & SR$\uparrow$ & MDD\%$\downarrow$ & CR\%$\uparrow$ & SR$\uparrow$ & MDD\%$\downarrow$ \\
    \midrule
    \multirow{6}{*}{\begin{tabular}[c]{@{}c@{}}Rule-\\based\end{tabular}} 

    & 	B\&H	& 122.77  & 0.94 & 25.12 & 88.75  & 0.86 & 26.28 \\
    & 	CRP	& 99.71 & 0.68  & 25.24  & 74.47  & 0.66  & 26.18\\
    & 	EG	& 99.82  & 0.68  & 25.41  & 74.98  & 0.65 & 26.18 \\
    & 	UP	& 99.90  & 0.68  & 25.36 & 73.45  & 0.64  & 26.68 \\
    & 	OLMAR	& -68.11  & -0.36  & 72.49 & -52.18  & -0.11 & 80.03 \\
    & 	WMAMR	& -52.59  & -0.23 & 60.50  & -71.60  & -0.35 & 86.22 \\
    \midrule

    \multirow{6}{*}{\begin{tabular}[c]{@{}c@{}}RL-\\based\end{tabular}} 
    & 	A2C	& 124.32 $\pm$ 9.21 & 0.95 $\pm$ 0.04 & 24.72 $\pm$ 1.11 & 92.42 $\pm$ 19.37  & 0.89 $\pm$ 0.13  & 23.43 $\pm$ 6.61 \\
    & 	PPO	& 122.55 $\pm$ 2.26 & 0.94 $\pm$ 0.01 & 25.24 $\pm$ 0.29 & 88.44 $\pm$ 6.11  & 0.87 $\pm$ 0.03  & 25.92 $\pm$ 1.65 \\
    & 	SAC & 120.63 $\pm$ 6.46 & 0.92 $\pm$ 0.02 & 25.47 $\pm$ 1.04 & 83.06 $\pm$ 14.76 & 0.82 $\pm$ 0.11  & 27.50 $\pm$ 3.72 \\
    & 	EIIE &	123.08 $\pm$ 0.73 & 0.94 $\pm$ 0.00 & 25.10 $\pm$ 0.12 & 87.83 $\pm$ 6.72 & 0.86 $\pm$ 0.05 & 26.15 $\pm$ 1.65\\
    & 	SARL	& 116.08 $\pm$ 2.72 & 0.91 $\pm$ 0.00 &  25.00 $\pm$ 0.19 & 79.22 $\pm$ 5.15  & 0.81 $\pm$ 0.05 & 26.63 $\pm$ 3.18 \\
    & 	Cross-Insight & 127.70 $\pm$ 9.03 & 0.94 $\pm$ 0.05 & 26.41 $\pm$ 1.41 & 97.13 $\pm$ 16.11 & 0.83 $\pm$ 0.10 & 28.15 $\pm$ 2.57 \\
    & AlphaGAT & $122.59 \pm 0.00$ & $0.93 \pm 0.00$ & $25.09 \pm 0.00$ & $87.09 \pm 0.15$ & $0.86 \pm 0.00$ & $25.86 \pm 0.00$ \\
    \cmidrule{2-8}
    & \textbf{\texttt{\ourm{}}} & $\mathbf{133.29 \pm 5.26}$ & $\mathbf{1.00 \pm 0.02}$ & $\mathbf{23.69 \pm 0.69}$ & $\mathbf{97.80 \pm 2.00}$ & $\mathbf{0.95 \pm 0.02}$ & $\mathbf{22.88 \pm 0.51}$ \\
    \bottomrule
    \end{tabular}}}
    \label{table:other_comparison}
\end{table*}

We conduct our experiments on five datasets covering diverse financial markets. 
For US equity markets, we use three widely adopted benchmarks: DOW30, NAS100, and SP500.
We also introduce the Japanese market dataset NIKKEI30, which contains 29 Nikkei component stocks, and the commodity ETF dataset COMMODITY\_ETF, which includes 7 mainstream commodity ETFs (Gold, Silver, Combined Commodity Index, Crude Oil, Natural Gas, Agricultural, and S\&P GSCI).
The raw data is obtained from Yahoo Finance, and we select constituent stocks with continuous listing and data availability throughout the entire period, resulting in 29, 73, 398, 29, and 7 assets for DOW30, NAS100, SP500, NIKKEI30, and COMMODITY\_ETF, respectively.
The daily data spans from May 1, 2008, to April 29, 2025, covering a total of 17 years that encompass various market regimes, including bull and bear markets, financial crises, and periods of high volatility.
The first 12 years are designated as the offline training set, while the subsequent 5 years are reserved for online evaluation.
The statistics of the datasets are summarized in Table \ref{tab:datasets}.

For feature engineering, the raw data includes daily open, high, low, close prices, and trading volume for each asset.
We compute adjusted prices, Moving Average Convergence Divergence (MACD), Bollinger Bands, and 17 common technical indicators to enrich the feature set.
Additional features include turbulence and VIX, which are incorporated to facilitate regime detection.
To avoid look-ahead bias and ensure realistic evaluation, all pre-processing steps, including feature engineering and normalization, are performed in a strictly time-series fashion, using only past and current information at each time step.
For reproducibility, the asset-level trading state uses 26 features per asset: open, high, low, close, volume, MACD, Bollinger upper and lower bands, RSI-30, CCI-30, DX-30, 30-day and 60-day moving averages, adjusted-price returns over 5/10/15/20/25/30 days, normalized open/high/low prices, adjusted-price return, close return, VIX, and turbulence.
The market-level input used by ARD and the RGM consists of six regime-sensitive signals: VIX, turbulence, Bollinger upper and lower bands, the 5-day adjusted-price return, and RSI-30.

\paragraph{Evaluation Metrics.}
To comprehensively evaluate the performance and robustness of PM methods, we employ three of the widely accepted financial metrics, including \textit{Cumulative Return} (CR), \textit{Sharpe Ratio} (SR), and \textit{Maximum DrawDown} (MDD).
These metrics capture different aspects of portfolio performance, including profit, risk-adjusted profit, and risk. 
They are computed online over the entire test period unless otherwise specified. 
Moreover, we follow the standard practice of CL \citep{wolczyk2021continual,pan2025multigranularity,sam2022cora} and use three metrics based on the agent's profit throughout different phases of its trading process: \textit{Average Performance} (AP), \textit{Forgetting} (FG), and \textit{Forward Transfer} (FT) \citep{pan2025survey}.

\paragraph{Baselines.}
To ensure a comprehensive evaluation, we select a range of representative methods, including six rule-based and seven RL-based methods.
Rule-based methods include Buy and Hold (B\&H), CRP \cite{helmbold1998line}, EG \cite{helmbold1998line}, UP \cite{blum1997universal}, OLMAR \cite{li2012online}, and WMAMR \cite{li2013weighted}, which are implemented based on PGPortfolio \cite{jiang2017deep}.
RL-based methods include A2C \cite{mnih2016asynchronous}, PPO \cite{schulman2017proximal}, SAC \cite{haarnoja2018soft}, EIIE \cite{jiang2017deep}, SARL \cite{ye2020reinforcement}, Cross-Insight \cite{zheng2024cross}, and AlphaGAT \cite{li2025AlphaGATTwoStage}, which are implemented based on FinRL \cite{liu2021finrl}.
Among them, EIIE is the first work formulating PM as an MDP with parallel independent evaluators; SARL augments the state with price movement predictions; Cross-Insight integrates multi-horizon investment insights; and AlphaGAT employs a two-stage framework with a CATimeMixer network for alpha factor mining.
For CL strategies, we compare with rolling-window Retraining (Retrain), continuous Finetuning (Finetune), Experience Replay (ER) \cite{isele2018selective} with a buffer size of 3,000, EWC-regularized finetuning \cite{james2017overcoming} with a regularization coefficient of 1,000, and Constrained Rationals (CoR) \cite{surdej2025BalancingExpressivity}, all using PPO as the underlying RL algorithm.
The Retrain window size is set to 12 years, matching the training period.

\paragraph{Implementation Details.}
We implement our framework via PyTorch and FinRL \cite{liu2021finrl}.
For each RL-based method, we use 10 different random seeds and report the average results and standard deviations to account for stochasticity.
We match the total RL update budget across RL-based methods, setting the training steps on each task to $10^4$ for methods that require task segmentation.
The actor and critic networks each consist of 2 layers of MLP with Tanh activation function, with an embedding dimension of 64.
The input state has dimensions of $(B, F, N)$, where $B=256$ is the batch size, $F=26$ is the number of features per asset, and $N$ is the number of assets.
We use z-score normalization to standardize features for training stability.
All methods use Adam optimizer with a learning rate of $10^{-4}$, and transaction costs are set to 10 basis points per trade.
For SARL, EIIE, AlphaGAT and \ourm{}, PPO is used as the underlying RL algorithm, while Cross-Insight uses its official SAC-based implementation.
For \ourm{}, we set $\kappa$ and $h$ to 0.5 and 2.5 times the standard deviation based on statistical experience, respectively.
The threshold for merging similar policy vectors $\delta_{\text{s}}$ is set to 0.5.
Following recent work in continual RL, policy vector learning and reuse are applied only to the actor, while the critic is reinitialized at each regime.
\footnote{Code available at: https://github.com/Dumail/ReCAP}

\subsection{Comparative Analysis}
To comprehensively evaluate the effectiveness and adaptability of our proposed framework, we conduct extensive comparative experiments against a diverse set of baseline methods. 
Our analysis is structured along two main axes: (1) comparison with established PM algorithms, including both traditional and RL-based methods, and (2) comparison with different CL strategies within the RL-based PM paradigm.

\subsubsection{Comparison with PM Methods.}
Table \ref{table:comparison} shows the performance of \ourm{} and other PM methods on the three datasets.
All methods are trained on the entire training period and evaluated on the test period, while \ourm{} is also continuously trained on the test period.
To ensure a fair comparison, the total training budget of \ourm{} is matched to that of the other RL-based PM methods.
We further report matched rolling-window adaptation results for the main PM baselines in Appendix \ref{app: more_experimental_results}.
Under a matched adaptive protocol on NAS100, using a 360-day window, a 90-day retraining frequency, a 180-day minimum window, and $10^4$ update steps per stage, the strongest adaptive baseline reaches 124.24\% CR and 0.92 SR, while \ourm{} achieves 164.89\% CR and 1.14 SR.

Among rule-based methods, CRP, EG, and UP achieve moderate returns but are generally limited by their inability to adapt to changing market conditions, resulting in higher drawdowns and lower risk-adjusted returns. 
Notably, B\&H's portfolio values align with market trends, as it evenly distributes capital across assets initially and sells them all at the end of the period.
Mean reversion-based methods (OLMAR and WMAMR) perform significantly worse, often suffering large losses during long-term trading, highlighting the vulnerability of static heuristics in non-stationary regimes.

\begin{figure}
    \centering
    \includegraphics[width=0.47\textwidth]{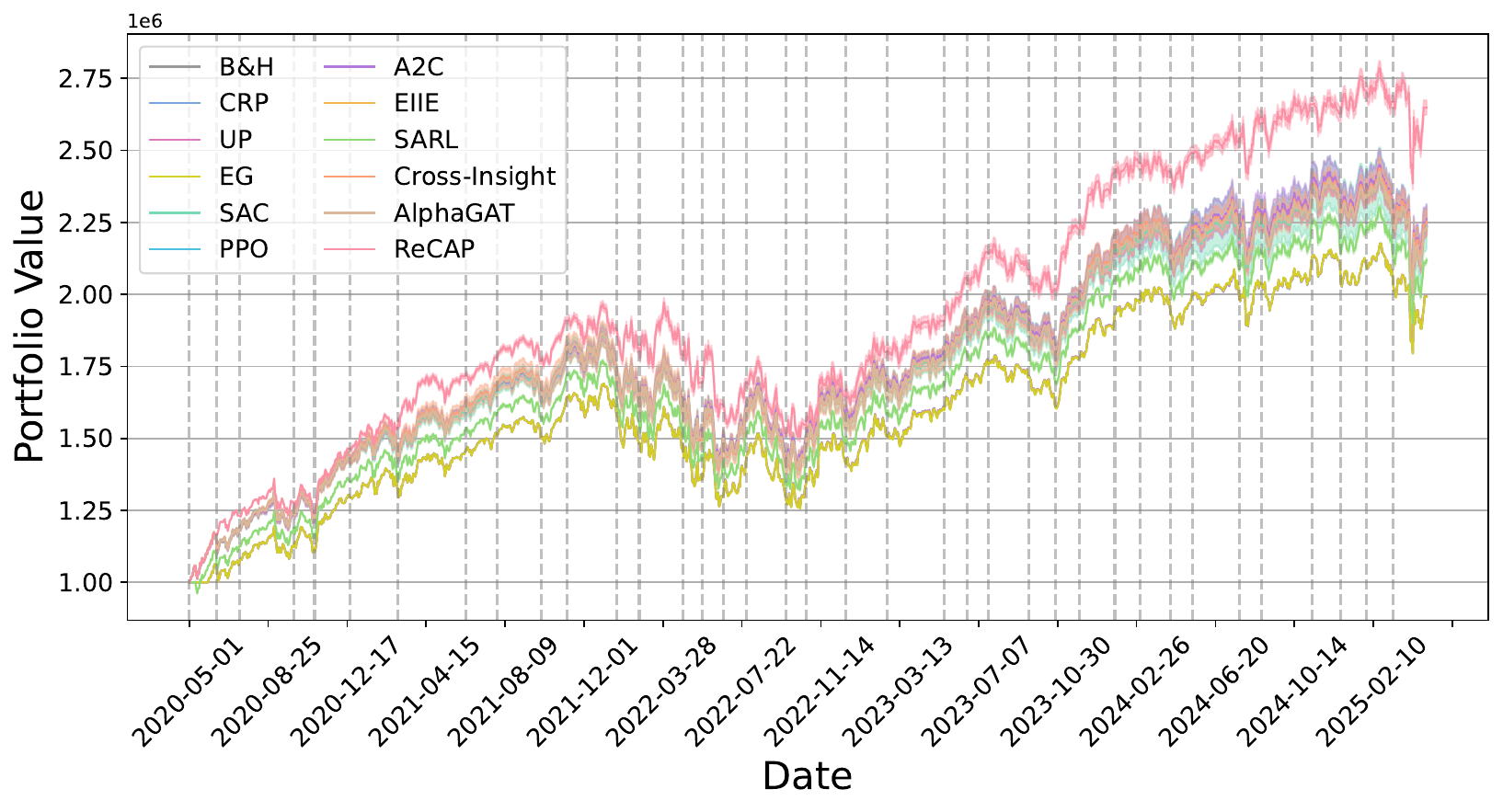}
    \caption{The portfolio values achieved by \ourm{} and other PM methods on the NAS100 dataset. 
    OLMAR and WMAMR are discarded due to their poor performance.
    The solid lines represent the mean portfolio values of each method, while the shaded areas indicate the standard deviation of the portfolio values.
    The gray dashed line represents the start date of each regime in \ourm{}.
    }
    \label{fig: values}
\end{figure}

RL-based methods demonstrate improved performance over most rule-based strategies. 
Among RL baselines, PPO and SAC achieve competitive results, with PPO serving as a strong baseline with low standard deviation in most cases.
In contrast, AlphaGAT exhibits poor performance, likely due to its complex two-stage architecture involving a CATimeMixer network. 
The significant imbalance between its large parameter space and the limited sample size leads to severe overfitting.
Cross-Insight, which integrates multiple investment insights across different horizons, achieves better performance than most RL baselines by effectively capturing multi-scale market dynamics.
Despite some advancements, most RL methods still assume stationary environments and lack explicit mechanisms for continual adaptation, which limits their ability to capitalize on regime shifts and recurring patterns. 
In contrast, our \ourm{} framework achieves the best results across all datasets, with an average cumulative return improvement of over $6.45\%-38.79\%$ compared to the strongest baselines. 
The lowest maximum drawdown of \ourm{} further indicates enhanced risk management and robustness to market turbulence. 

To verify the generalization capabilities of our proposed framework across diverse financial markets, we further conduct experiments on the NIKKEI30 and COMMODITY\_ETF datasets, with results presented in Table \ref{tab:comparison_new}. 
Consistent with observations from the US equity markets, \ourm{} demonstrates superior performance on both the Japanese equity market and the commodity ETF market.
Specifically, on the NIKKEI30 dataset, \ourm{} achieves a Cumulative Return (CR) of 133.29\% and a Sharpe Ratio (SR) of 1.00, surpassing the second-best baseline, Cross-Insight, which yields a CR of 127.70\%. 
In the COMMODITY\_ETF dataset, which exhibits distinct volatility patterns compared to equities, \ourm{} maintains its dominance.
These findings underscore the robustness of \ourm{} and its ability to effectively adapt to the unique dynamics of different asset classes and market environments.

To visualize the performance differences, Figure \ref{fig: values} shows the portfolio values achieved by \ourm{} and other PM methods across the NAS100.
Results on other datasets are provided in the Appendix \ref{app: more_experimental_results}.
From the figure, it is evident that our framework consistently outperforms other models, regardless of the bull or bear market.
Specifically, \ourm{} segments the test period into multiple variable-length regimes and adapts its strategy accordingly, resulting in superior performance.
Furthermore, as the trading duration increases, our method gradually accumulates knowledge of different regimes through the Policy Library, making the gap between it and other methods more pronounced and allowing it to achieve greater excess returns.
This demonstrates the advantage of \ourm{} in long-term trading scenarios.

\subsubsection{Comparison with CL Strategies.}
To specifically investigate the benefits of \ourm{}, we compare with five CL strategies based on PPO.  
Other details are provided in the Appendix \ref{app: more_experimental_results}.
The task boundaries for these baselines are determined using fixed-length windows to match common practice in the literature.
We additionally report quarterly and yearly exogenous task boundaries in Appendix \ref{app: more_experimental_results}, which lead to the same qualitative conclusion.

The results in Table \ref{table:cl} provide a comprehensive comparison in terms of CL metrics.
Across three US market datasets, our proposed \ourm{} framework consistently achieves the highest average performance by a substantial margin, outperforming all other CL baselines.
This demonstrates the strong adaptability and knowledge accumulation capability of \ourm{} when facing long-term, non-stationary market environments.
Furthermore, as the number of assets increases, the performance gap widens.
On the SP500, \ourm{} attains a nearly double average performance that of the best baseline. 
This indicates that our framework is particularly effective in complex, high-dimensional environments.

In terms of forgetting and forward transfer, which quantify the retention of knowledge and the ability to leverage prior knowledge, \ourm{} achieves comparable or superior results relative to other CL strategies. 
Notably, with the increase in the number of assets, the forgetting and forward transfer values of all methods approach zero.
This may be due to the increased independence between tasks when more assets are involved.
Other widely used CL strategies, including Finetune, ER, and EWC, do not show obvious improvements over the retrain strategy. 
Moreover, ER and EWC represent replay-based and regularization-based CL strategies, respectively, and their overall performance is not superior to native Finetune.
CoR attempts to balance plasticity and stability through constraint optimization. 
However, as shown in Table \ref{table:cl}, its performance in the financial domain remains to be verified.
This may be because task segmentation in online trading does not align well with static data distributions, further highlighting the unique challenges of continual PM.

\begin{table}
\renewcommand{\arraystretch}{0.7}
\setlength{\abovecaptionskip}{0.1cm}
\centering
\caption{Performance comparison of five CL strategies and \ourm{} on three datasets.
``Static'' refers to training on the entire training period and evaluating during the test period.
}
\begin{threeparttable}
\footnotesize
\setlength{\tabcolsep}{1mm}{
\resizebox{0.89\linewidth}{!}{
\begin{tabular}{cc|cccc} 
\toprule

\begin{tabular}[c]{@{}c@{}}Dataset\end{tabular} & \begin{tabular}[c]{@{}c@{}}Strategy\end{tabular} & AP\%$\uparrow$ & FT\%$\uparrow$  & FG\%$\downarrow$ \\ 

\midrule

\multirow{4}{*}{\begin{tabular}[c]{@{}c@{}}DOW30\end{tabular}} 
& Static  & 13.07 $\pm$ 0.19 & -  & -  \\
& Retrain & 18.26 $\pm$ 0.08 & -  & -  \\
& Finetune & 18.34 $\pm$ 0.25 & \textbf{2.82 $\pm$ 11.39}  & \textbf{-1.88 $\pm$ 5.67}  \\
& ER & 18.47 $\pm$ 0.47 & 0.55 $\pm$ 26.41 & 1.04 $\pm$ 21.14 \\
& EWC & 18.35 $\pm$ 0.18 & -1.90 $\pm$ 4.11 & 0.78 $\pm$ 1.75  \\
& CoR & $18.31 \pm 0.30$ & $1.60\pm10.02$ &  $3.89\pm10.39$ \\
\cmidrule{2-5}
& \textbf{\texttt{\ourm{}}} & \textbf{24.92 $\pm$ 0.65} & 1.35 $\pm$ 3.83 & -0.69 $\pm$ 1.92  \\

\midrule

\multirow{4}{*}{\begin{tabular}[c]{@{}c@{}}NAS100\end{tabular}} 
& Static  & 17.53 $\pm$ 0.14 & -  & -  \\
& Retrain & 24.51 $\pm$ 0.05 & -  & -  \\
& Finetune & 24.59 $\pm$ 0.18 & 0.03 $\pm$ 0.23  & \textbf{-0.06 $\pm$ 0.08}  \\
& ER & 24.49 $\pm$ 0.24 & -0.02 $\pm$ 0.35 & -0.02 $\pm$ 0.14 \\
& EWC & 24.56 $\pm$ 0.15 & 0.01 $\pm$ 0.02  & -0.01 $\pm$ 0.01  \\
& CoR & $24.53 \pm 0.09$ & $0.01\pm0.00$ &  $0.01\pm0.01$\\
\cmidrule{2-5}
& \textbf{\texttt{\ourm{}}} & \textbf{39.03 $\pm$ 0.35} & \textbf{0.03 $\pm$ 0.07} & -0.02 $\pm$ 0.03  \\

\midrule

\multirow{4}{*}{\begin{tabular}[c]{@{}c@{}}SP500\end{tabular}} 
& Static  & 16.18 $\pm$ 0.03 & -  & -  \\
& Retrain & 22.86 $\pm$ 0.01 & -  & -  \\
& Finetune & 22.85 $\pm$ 0.04 & 0.01 $\pm$ 0.05  & -0.01 $\pm$ 0.03  \\
& ER & 22.85 $\pm$ 0.07 & \textbf{0.01 $\pm$ 0.04} & \textbf{-0.01 $\pm$ 0.02} \\
& EWC & 22.87 $\pm$ 0.03 & 0.00 $\pm$ 0.01  & 0.00 $\pm$ 0.00  \\
& CoR & $22.87 \pm 0.03$ & $0.00\pm0.00$ & $0.00\pm0.00$ \\
\cmidrule{2-5}
& \textbf{\texttt{\ourm{}}} & \textbf{47.18 $\pm$ 0.43} & 0.00 $\pm$ 0.09 & -0.00 $\pm$ 0.05  \\

\bottomrule

\end{tabular}
}}
\end{threeparttable}
\label{table:cl}
\end{table}

\subsection{Ablation Studies}
To further understand the contribution of each component in our proposed \ourm{} framework, we conduct ablation studies on the NAS100. 
We compare the full model with three ablated variants: 
1) \textbf{w/o-ARD}, where the market is segmented using fixed-length windows instead of using ARD; 
2) \textbf{w/o-PL}, where only the most recent policy vector is retained and reused, discarding the accumulated policy library;
and 3) \textbf{w/o-RGM}, where the RGM is removed and policy vectors are combined using random weights; 

\begin{figure}
    \centering
    \includegraphics[width=0.4\textwidth]{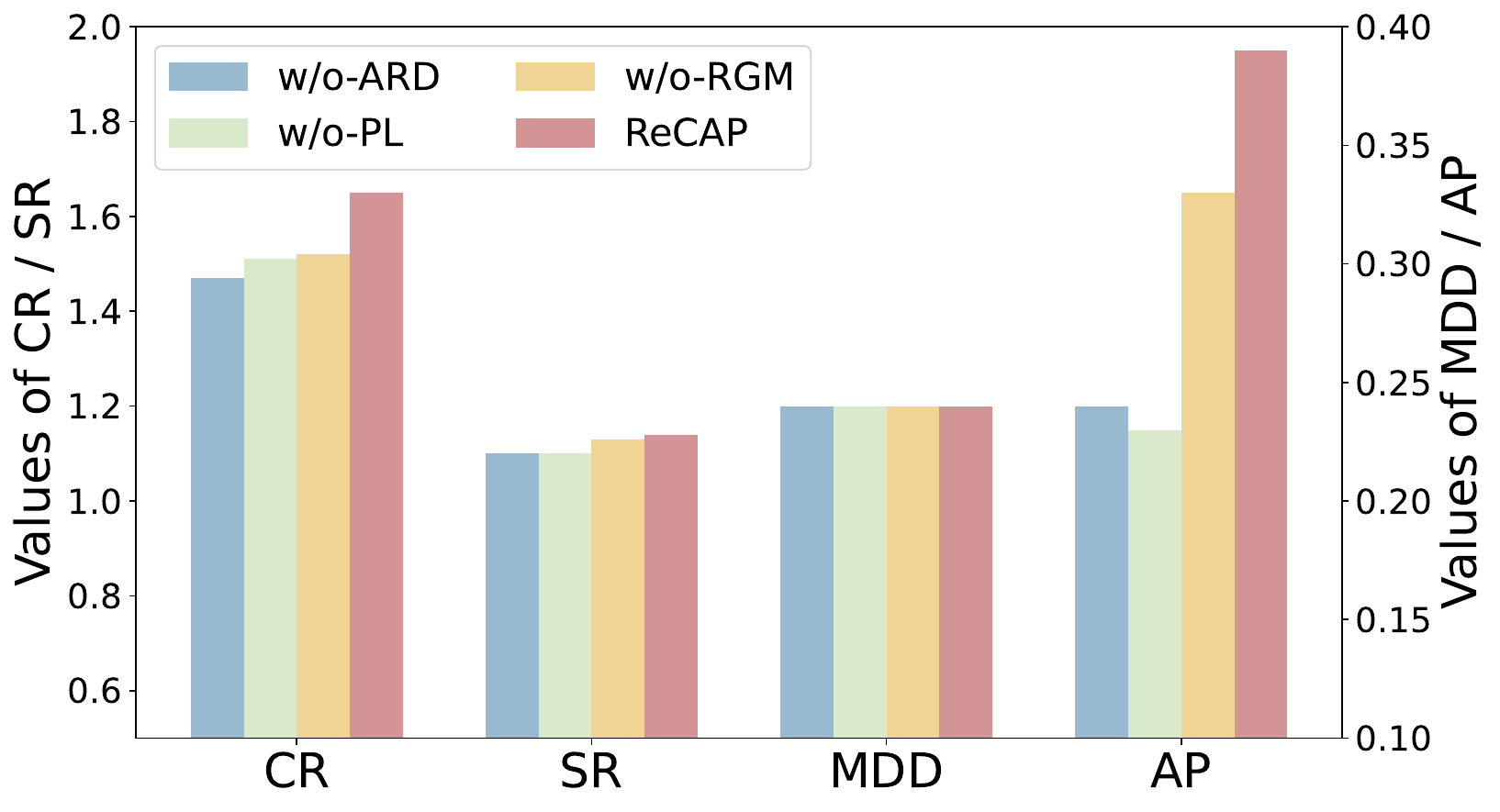}
    \caption{Ablation study of \ourm{} on the NAS100.
We compare \ourm{} with three ablated variants: without adaptive regime detection (w/o-ARD), without policy library (w/o-PL), and without regime-gate module (w/o-RGM).
}
    \label{fig: ablation}
\end{figure}

Figure~\ref{fig: ablation} summarizes the results across four evaluation metrics.
Among the ablated variants, w/o-ARD exhibits the largest drop in performance across all indicators.
This result underscores the critical importance of accurate and adaptive regime segmentation for continual PM. 
Without proper regime identification, the agent is unable to align its learning and adaptation process with true market dynamics. 
The w/o-PL variant performs second worst, indicating that the absence of a policy library severely limits the agent's ability to accumulate and reuse knowledge from past regimes. 
This also leads to the worst average performance of the final policy on all regimes, i.e., the lowest AP.
The full framework consistently achieves the best performance on all metrics, demonstrating that all three components are indispensable for addressing continual PM. 
The synergy among these modules enables \ourm{} to not only accumulate and retain valuable knowledge but also to deploy it effectively in response to evolving market conditions.

\subsection{Parameter Sensitivity}
We further investigate the sensitivity of \ourm{} to the hyperparameters in the adaptive regime detection module, specifically the drift parameter $\kappa$ and the regime change threshold $h$.
Table \ref{tab:parameter} presents the performance on NAS100 under different parameter settings, where $\kappa$ and $h$ are set as multiples of the standard deviation $\sigma$.
The results show that \ourm{} maintains superior performance across a range of parameter values.
Varying $\kappa$ while fixing $h$ consistently yields high cumulative returns and Sharpe ratios.
Similarly, the performance remains robust when adjusting $h$.
This indicates that the effectiveness of \ourm{} is not sensitive to precise hyperparameter tuning, demonstrating the robustness of our proposed framework.
Replacing CUSUM with an HMM equipped with BIC-based model selection on NAS100 still yields 146.38\% CR, 1.10 SR, and 24.22\% MDD, indicating that the gain does not depend on a single detector implementation.
Under additional +5 bps and +10 bps cost settings, \ourm{} preserves a clear advantage. Detailed results are provided in Appendix \ref{app: more_experimental_results}.

\subsection{More Discussion}
The experimental results show that \ourm{} outperforms both popular PM methods and common CL strategies.
Two observations are important. First, task segmentation that better matches true regime shifts is important for continual adaptation, as fixed-window segmentation leads to noticeable degradation. Second, the combination of adaptive regime detection, policy-vector accumulation, and regime-aware policy composition improves long-horizon return, risk-adjusted performance, and drawdown control in a unified framework. The modular design also leaves room for stronger regime detectors, knowledge distillation, and multi-agent extensions. 
Limitations and future directions are discussed in the Appendix \ref{app: limitations_future_work}.

\begin{table}[h]
\centering
\renewcommand{\arraystretch}{0.7}
\caption{Parameter sensitivity analysis on NAS100.
We compare different settings of the drift parameter $\kappa$ and the regime change threshold $h$.
}
\label{tab:parameter}
\resizebox{0.4\textwidth}{!}{
\begin{tabular}{ccccc}
\toprule
$\kappa$ & $h$ & CR\%$\uparrow$ & SR$\uparrow$ & MDD\%$\downarrow$ \\
\midrule
$0.3\sigma$ & $2.5\sigma$ & 164.12 $\pm$ 3.18 & 1.15 $\pm$ 0.02 & 17.07 $\pm$ 0.72 \\
$0.5\sigma$ & $2.5\sigma$ & 164.89 $\pm$ 2.54 & 1.14 $\pm$ 0.01 & 23.86 $\pm$ 0.35 \\
$0.7\sigma$ & $2.5\sigma$ & 172.07 $\pm$ 3.80 & 1.17 $\pm$ 0.03 & 17.03 $\pm$ 0.70 \\
$0.5\sigma$ & $2.0\sigma$ & 160.21 $\pm$ 0.05 & 1.12 $\pm$ 0.00 & 25.82 $\pm$ 0.11 \\
$0.5\sigma$ & $3.0\sigma$ & 132.61 $\pm$ 0.00 & 1.00 $\pm$ 0.00 & 26.60 $\pm$ 0.13 \\
\bottomrule
\end{tabular}
}
\end{table}

\section{Conclusion}
\label{sec: conclusion}

In this work, we identified and attempted to address the fundamental challenge of non-stationarity in financial markets, a pervasive issue due to ``alpha decay''. 
We proposed \textbf{\ourm{}} (\textbf{Re}gime-aware \textbf{C}ontinual \textbf{A}daptive \textbf{P}ortfolio management), a principled framework that reformulates portfolio management as a continual learning process. 
\ourm{} introduces a novel synergy between adaptive regime detection, a reusable policy library, and a regime-gate mechanism, enabling the agent to dynamically detect market shifts, retain historical expertise, and rapidly adapt to novel regimes.
Comprehensive experiments on five real-world datasets, covering major US and Japanese stock indices as well as commodity ETFs, demonstrate the superiority of our approach. 
\ourm{} not only consistently outperforms state-of-the-art rule-based and RL-based baselines in terms of profitability and risk control but also surpasses existing continual learning strategies in average performance and forward transfer. 
These results highlight the effectiveness of \ourm{} in balancing continual adaptation and knowledge retention in dynamic financial environments. 

\section{Acknowledgments}
This work was supported by the National Natural Science Foundation of China (Nos. 62476228, 62506308), and the Chengdu Science and Technology Program (No.2025-YF12-00030-RC).

\bibliographystyle{ACM-Reference-Format}
\balance
\bibliography{main}

\appendix
\section*{Appendix} 
\section{Definitions}
\label{app: definitions}

\begin{definition}[OHLCV]
\label{def: ohlcv}
The Open-High-Low-Close-Volume is a time series representation of asset prices.
For a market consisting of $N$ tradable assets, let $\mathbf{X}_t = \{x_{t,1}, \ldots, x_{t,N}\} $ denote the market observation at time $t$. 
Each asset $i \in [1,N]$ is represented by OHLCV, that is $x_{t,i} = \left[ O_{t,i}, H_{t,i}, L_{t,i}, C_{t,i}, V_{t,i} \right]$, where $O_{t,i}$, $H_{t,i}$, $L_{t,i}$, $C_{t,i}$, and $V_{t,i}$ denote the open, high, low, close prices, and trading volume of asset $i$ at time $t$, respectively.
\end{definition}

\begin{definition}[Technical Indicators]
\label{def: technical indicators}
Technical indicators are derived from OHLCV data and are used to analyze market trends and volatility.
Typical indicators include Moving Average (MA), Relative Strength Index (RSI), and Moving Average Convergence Divergence (MACD).
We use $\mathbf{f}_{t, i}$ to denote the concatenation of raw OHLCV $x_{t, i}$ and technical features for asset $i$ at time $t$.
\end{definition}

\begin{definition}[Portfolio]
\label{def: portfolio}
A portfolio is a collection of assets held by an investor.
At time $t$, it is defined by a vector of weights $\mathbf{w}_t = [w_{t,0},w_{t,1}, \ldots, w_{t,N}]$, where $w_{t,i}$ represents the proportion of total capital allocated to asset $i\in[1,N]$ or risk-free cash ($i=0$), subject to the constraints $w_{t,i} \geq 0, \sum_{i=0}^{N} w_{t,i} = 1 $.
Short selling is not considered in this work, but can be incorporated by relaxing the non-negativity constraint.
\end{definition}

\begin{definition}[Portfolio Value and Return]
\label{def: portfolio value}
The portfolio value at time $t$ is defined as the total market value of all assets held, including the cash. 
Let $V_t$ denote the portfolio value at time $t$. 
Assuming an initial capital $V_0$, the portfolio value evolves as follows:
\begin{equation}
    V_t = V_{t-1} \cdot \left( \mathbf{w}_{t-1}^\top \mathbf{r}_t \right) \cdot (1 - c_t),
\end{equation}
where $\mathbf{w}_{t-1}$ is the portfolio at time $t-1$, $\mathbf{r}_t = [1, r_{t, 1}, \ldots, r_{t, N}]^\top$ is the vector of gross returns for each asset (with $r_{t, i} = \frac{C_{t, i}}{C_{t-1, i}}$), and $c_t$ denotes the proportional transaction cost incurred at time $t$.
Then, the single-period portfolio return at time $t$ is defined as the relative change in portfolio value:
\begin{equation}
    R_t = \frac{V_t - V_{t-1}}{V_{t-1}}.
\end{equation}
\end{definition}

\begin{figure*}[ht]
    \centering
    \begin{subfigure}[b]{0.47\textwidth}
        \centering
        \includegraphics[width=\linewidth]{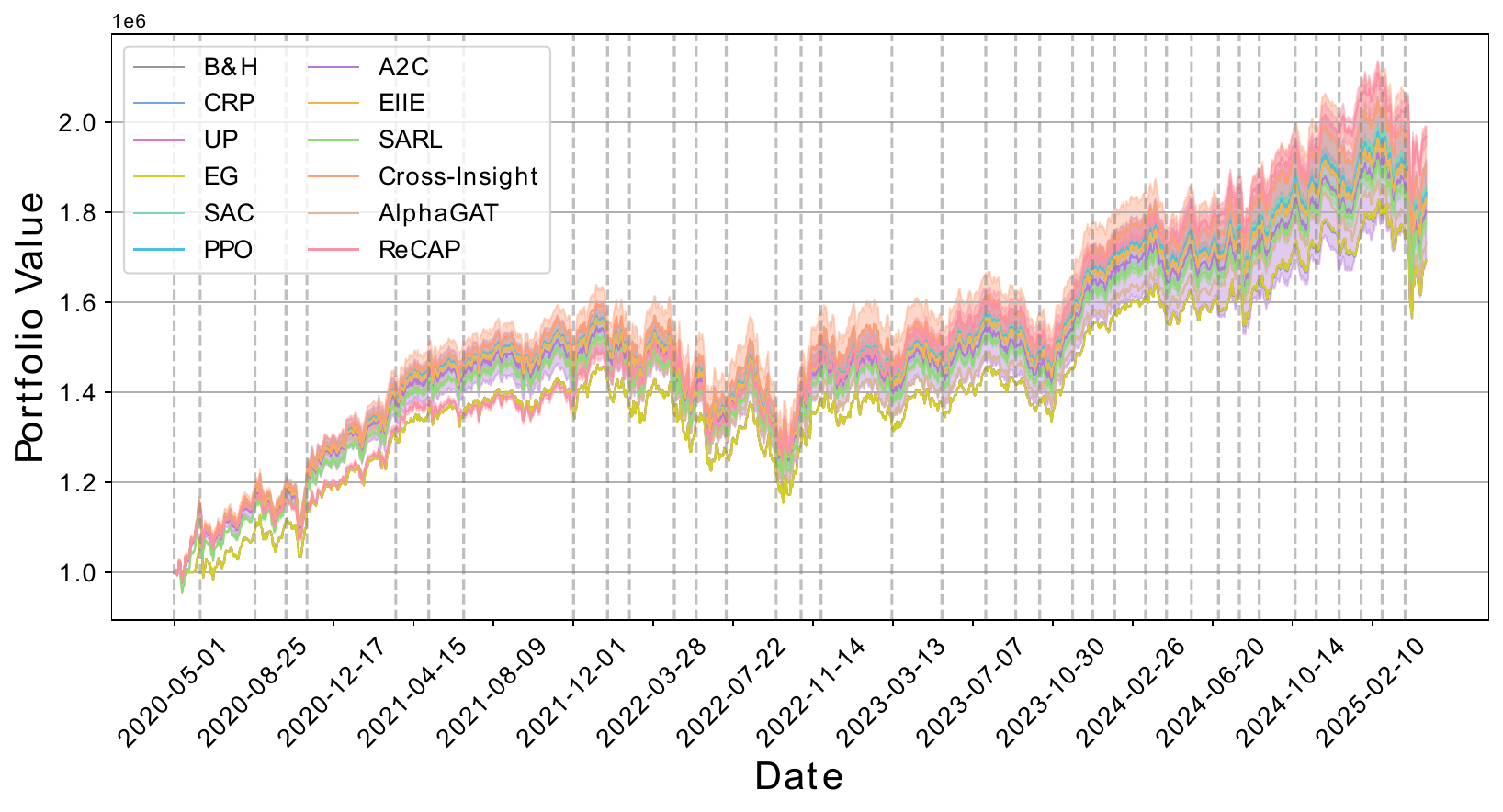}
        \caption{DOW30}
        \label{fig: dow30}
    \end{subfigure}
    \begin{subfigure}[b]{0.47\textwidth}
        \centering
        \includegraphics[width=\linewidth]{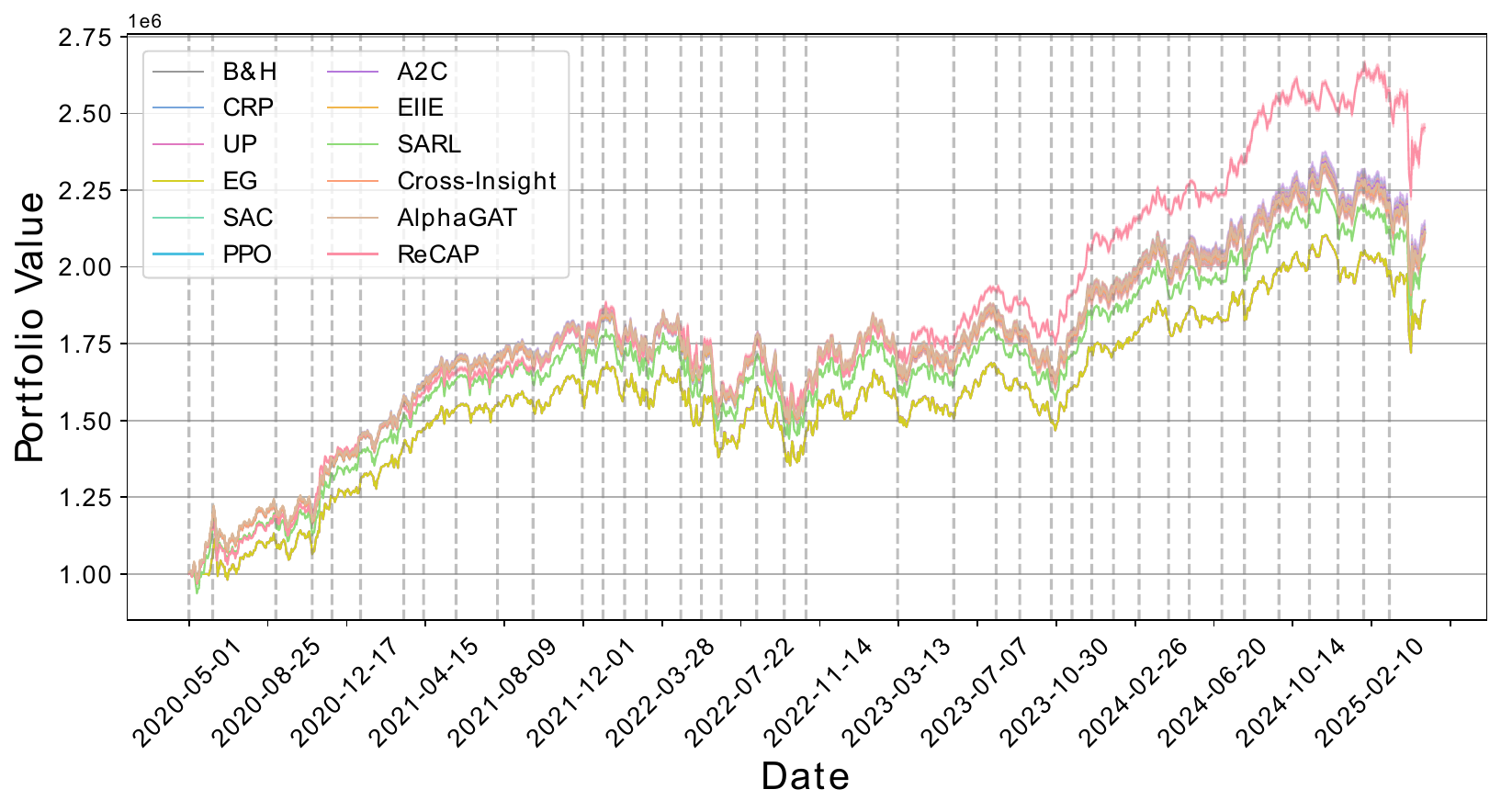}
        \caption{SP500}
        \label{fig: sp500}
    \end{subfigure}
    \\
    \begin{subfigure}[b]{0.47\textwidth}
        \centering
        \includegraphics[width=\linewidth]{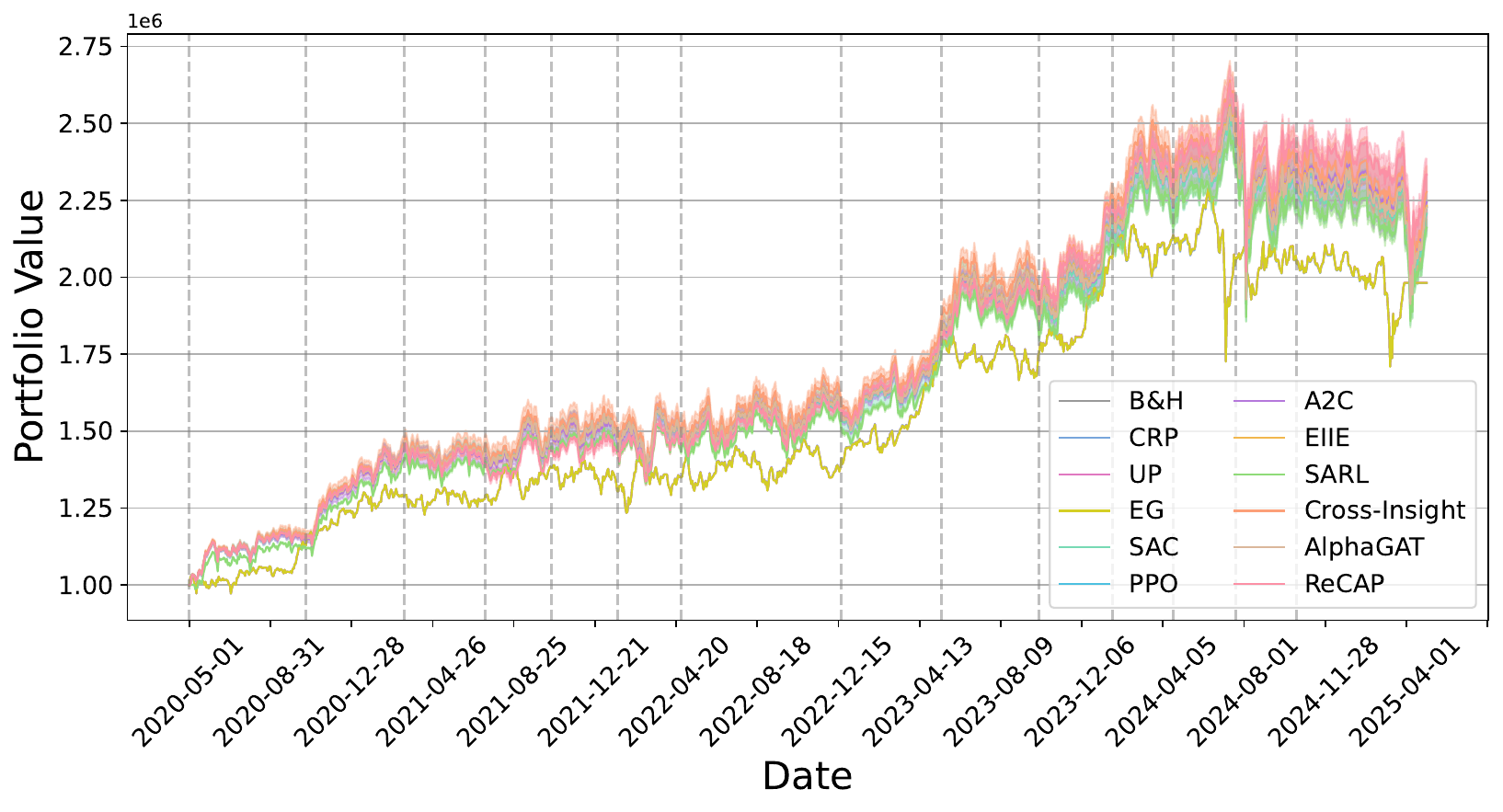}
        \caption{NIKKEI30}
        \label{fig: nikkei30}
    \end{subfigure}
    \begin{subfigure}[b]{0.47\textwidth}
        \centering
        \includegraphics[width=\linewidth]{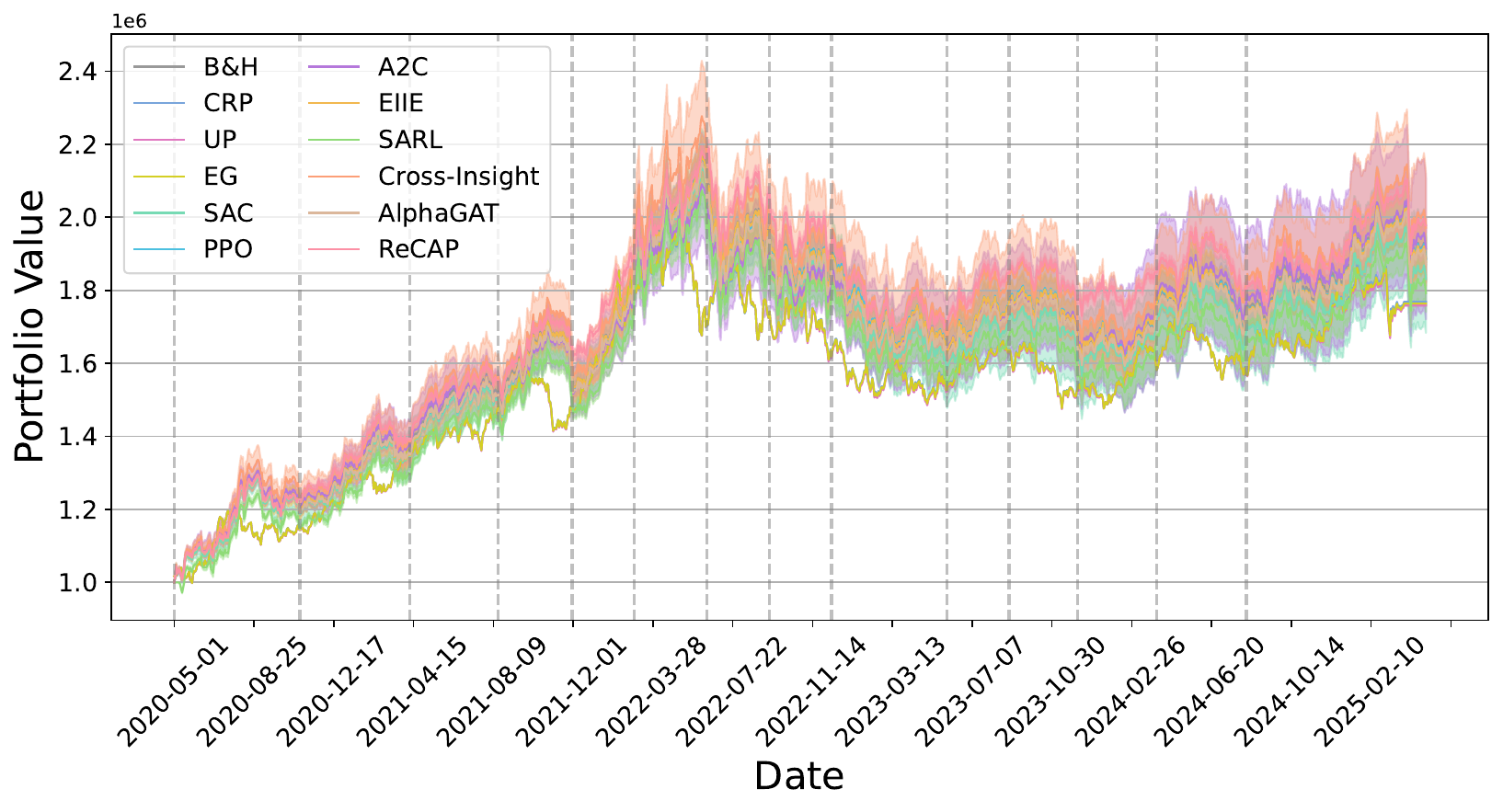}
        \caption{COMMODITY\_ETF}
        \label{fig: commodity_etf}
    \end{subfigure}
    \caption{The portfolio values achieved by \ourm{} and other PM methods on the DOW30, SP500, NIKKEI30 and COMMODITY\_ETF datasets. 
    OlMAR and WMAMR are discarded due to their poor performance.
    }
    \label{fig: more_values}
    
\end{figure*}

\begin{table}
\renewcommand{\arraystretch}{0.7}
\setlength{\abovecaptionskip}{0.1cm}
\centering
\caption{Performance comparison of five CL strategies and \ourm{} on NAS100 in terms of three PM metrics and AP.
}
\footnotesize
\setlength{\tabcolsep}{1mm}{
\resizebox{0.99\linewidth}{!}{
\begin{tabular}{c|cccc} 
\toprule

Method & CR\%$\uparrow$ & SR$\uparrow$ & MDD\%$\downarrow$ & AP\%$\uparrow$ \\
\midrule
Static & 123.54 $\pm$ 1.30 & 0.92 $\pm$ 0.01 & 25.37 $\pm$ 0.17 & 17.53 $\pm$ 0.14 \\
Pretrain & 145.83 $\pm$ 1.31 & 1.09 $\pm$ 0.01 & 24.57 $\pm$ 0.14 & 24.55 $\pm$ 0.12 \\
Retrain & 145.57 $\pm$ 0.45 & 1.09 $\pm$ 0.00 & 24.56 $\pm$ 0.04 & 24.51 $\pm$ 0.05 \\
Finetune & 145.60 $\pm$ 2.13 & 1.09 $\pm$ 0.01 & 24.53 $\pm$ 0.22 & 24.59 $\pm$ 0.18 \\
ER & 145.30 $\pm$ 1.49 & 1.09 $\pm$ 0.01 & 24.59 $\pm$ 0.23 & 24.49 $\pm$ 0.24 \\
EWC & 145.72 $\pm$ 1.37 & 1.09 $\pm$ 0.01 & 24.57 $\pm$ 0.18 & 24.56 $\pm$ 0.15 \\
\ourm{} & \textbf{164.89 $\pm$ 2.54} & \textbf{1.14 $\pm$ 0.01} & \textbf{23.86 $\pm$ 0.35} & \textbf{39.03 $\pm$ 0.35} \\

\bottomrule

\end{tabular}
}}
\label{table:cl_pm}
\end{table}

\section{More Experimental Results}\label{app: more_experimental_results}
Table \ref{table:cl_pm} compares \ourm{} with five CL strategies on NAS100 and also includes a Pretrain baseline that learns a base policy on the full training period before task-wise finetuning.
The Pretrain baseline performs similarly to Retrain, and all fixed-boundary CL variants improve over the Static baseline only within a limited range.
In contrast, \ourm{} achieves clearly stronger results across all metrics, indicating that adaptive regime detection and policy-vector accumulation are more effective than periodic reuse of data or parameters alone.

Figure \ref{fig: more_values} shows the portfolio values achieved by \ourm{} and other PM methods in the DOW30, SP500, NIKKEI30 and COMMODITY\_ETF datasets.
As the number of assets increases, the standard deviation of portfolio values decreases because the impact of individual asset fluctuations is averaged out.
\ourm{} demonstrates a more stable performance across different asset configurations.
On DOW30, \ourm{} initially underperforms, particularly between 2020-11-09 and 2021-12-30, likely due to the unique regime characteristics and longer regime duration during that period.
However, as knowledge accumulates, \ourm{} gradually surpasses other methods, achieving the highest portfolio value after 2024-04-25, outperforming the state-of-the-art method Cross-Insight.
On SP500, \ourm{} exhibits a similar trend. 
In the earlier period, it performs well but does not significantly exceed Cross-Insight.
However, after 2023-03-10, \ourm{} begins to diverge from other methods, ultimately achieving over 30\% excess return compared to them.
These observations highlight \ourm{}'s ability to leverage knowledge accumulation and adapt to changing market conditions effectively.
Moreover, \ourm{} demonstrates a characteristic of achieving greater excess returns in scenarios with a larger number of assets and longer investment horizons, aligning with the principles of CL and practical PM needs.

Table \ref{tab:fixed_boundary_cl} reports additional CL comparisons under exogenous quarterly and yearly task boundaries.
Even under these fixed schedules, the strongest baseline reaches 19.52\% AP on DOW30 and 28.45\% AP on NAS100, both remaining well below the corresponding ReCAP results in the main text.
This indicates that the advantage of \ourm{} does not depend on regime-derived task boundaries.

\begin{table*}[t]
\renewcommand{\arraystretch}{0.7}
\setlength{\abovecaptionskip}{0.1cm}
\centering
\caption{CL baselines under exogenous quarterly and yearly task boundaries.}
\footnotesize
\setlength{\tabcolsep}{1.2mm}{
\resizebox{0.98\linewidth}{!}{
\begin{tabular}{cc|ccc|ccc|ccc|ccc}
\toprule
& & \multicolumn{3}{c|}{Finetune} & \multicolumn{3}{c|}{EWC} & \multicolumn{3}{c|}{CoR} & \multicolumn{3}{c}{ER} \\
Dataset & Boundary & AP\%$\uparrow$ & FT\%$\uparrow$ & FG\%$\downarrow$ & AP\%$\uparrow$ & FT\%$\uparrow$ & FG\%$\downarrow$ & AP\%$\uparrow$ & FT\%$\uparrow$ & FG\%$\downarrow$ & AP\%$\uparrow$ & FT\%$\uparrow$ & FG\%$\downarrow$ \\
\midrule
DOW30 & Quarterly & 19.52 $\pm$ 0.40 & 0.95 $\pm$ 0.51 & 6.81 $\pm$ 14.05 & 19.43 $\pm$ 0.11 & 0.18 $\pm$ 0.32 & -0.10 $\pm$ 0.15 & 19.14 $\pm$ 0.28 & -1.30 $\pm$ 3.15 & 0.56 $\pm$ 4.03 & 19.39 $\pm$ 0.65 & 0.23 $\pm$ 0.30 & -9.08 $\pm$ 18.76 \\
DOW30 & Yearly & 14.41 $\pm$ 0.18 & 0.40 $\pm$ 0.82 & 0.19 $\pm$ 0.14 & 14.36 $\pm$ 0.19 & 0.04 $\pm$ 0.06 & -0.01 $\pm$ 0.04 & 14.35 $\pm$ 0.12 & -1.36 $\pm$ 0.75 & 0.82 $\pm$ 0.69 & 14.17 $\pm$ 0.19 & -0.22 $\pm$ 1.07 & 0.06 $\pm$ 0.80 \\
NAS100 & Quarterly & 28.22 $\pm$ 0.24 & 0.04 $\pm$ 0.11 & -0.03 $\pm$ 0.08 & 28.29 $\pm$ 0.12 & 0.01 $\pm$ 0.01 & -0.01 $\pm$ 0.01 & 28.45 $\pm$ 0.34 & 0.26 $\pm$ 0.10 & -0.16 $\pm$ 0.09 & 28.27 $\pm$ 0.27 & 0.09 $\pm$ 0.18 & -0.06 $\pm$ 0.16 \\
NAS100 & Yearly & 21.68 $\pm$ 0.16 & -0.05 $\pm$ 0.26 & 0.03 $\pm$ 0.10 & 21.70 $\pm$ 0.10 & 0.01 $\pm$ 0.04 & -0.00 $\pm$ 0.02 & 21.68 $\pm$ 0.17 & 0.17 $\pm$ 0.46 & -0.09 $\pm$ 0.21 & 21.77 $\pm$ 0.21 & -0.03 $\pm$ 0.17 & -0.06 $\pm$ 0.09 \\
\bottomrule
\end{tabular}}}
\label{tab:fixed_boundary_cl}
\end{table*}

Table \ref{tab:rolling_retrain_pm} reports a matched rolling-window adaptation comparison on NAS100, where competing PM baselines are also allowed to update online using a 360-day window, a 90-day retraining frequency, a 180-day minimum window, and $10^4$ update steps per retraining stage.
Under this matched adaptive protocol, \ourm{} still maintains a clear margin over all PM baselines.

\begin{table}[t]
\renewcommand{\arraystretch}{0.7}
\setlength{\abovecaptionskip}{0.1cm}
\centering
\caption{Matched rolling-window retraining comparison on NAS100.}
\footnotesize
\setlength{\tabcolsep}{1.5mm}{
\resizebox{0.89\linewidth}{!}{
\begin{tabular}{lccc}
\toprule
Method & CR\%$\uparrow$ & SR$\uparrow$ & MDD\%$\downarrow$ \\
\midrule
Cross-Insight & 124.24 $\pm$ 7.69 & 0.91 $\pm$ 0.03 & 26.08 $\pm$ 0.76 \\
PPO & 123.73 $\pm$ 1.16 & 0.92 $\pm$ 0.01 & 25.39 $\pm$ 0.15 \\
SAC & 123.55 $\pm$ 6.59 & 0.91 $\pm$ 0.03 & 25.59 $\pm$ 0.83 \\
A2C & 121.90 $\pm$ 7.12 & 0.90 $\pm$ 0.03 & 25.75 $\pm$ 0.36 \\
SARL & 112.31 $\pm$ 2.62 & 0.86 $\pm$ 0.01 & 25.31 $\pm$ 0.22 \\
AlphaGAT & 123.38 $\pm$ 2.30 & 0.92 $\pm$ 0.00 & 25.36 $\pm$ 0.45 \\
EIIE & 123.22 $\pm$ 0.15 & 0.92 $\pm$ 0.00 & 25.35 $\pm$ 0.05 \\
\ourm{} & \textbf{164.89 $\pm$ 2.54} & \textbf{1.14 $\pm$ 0.01} & \textbf{23.86 $\pm$ 0.35} \\
\bottomrule
\end{tabular}}}
\label{tab:rolling_retrain_pm}
\end{table}

To test robustness to the choice of detector, we replace CUSUM in ARD with an HMM equipped with BIC-based model selection on NAS100.
The resulting policy still achieves 146.38\% CR, 1.10 SR, and 24.22\% MDD over 5 seeds, confirming that the overall gain is not tied to a single detector implementation.

Table \ref{tab:cost_sensitivity} reports turnover and transaction-cost sensitivity on NAS100.
Although \ourm{} trades slightly more actively than A2C, it preserves a substantial advantage under both +5 bps and +10 bps cost settings.

\begin{table}[t]
\renewcommand{\arraystretch}{0.7}
\setlength{\abovecaptionskip}{0.1cm}
\centering
\caption{Turnover and transaction-cost sensitivity on NAS100.}
\footnotesize
\setlength{\tabcolsep}{1.5mm}{
\resizebox{0.89\linewidth}{!}{
\begin{tabular}{ccccc}
\toprule
Cost & Method & Turnover$\downarrow$ & CR\%$\uparrow$ & SR$\uparrow$ \\
\midrule
+5 bps & A2C & 0.0440 $\pm$ 0.0104 & 115.97 $\pm$ 0.03 & 0.87 $\pm$ 0.01 \\
+5 bps & \ourm{} & 0.0504 $\pm$ 0.0041 & \textbf{139.96 $\pm$ 2.88} & \textbf{1.067 $\pm$ 0.01} \\
+10 bps & A2C & 0.0403 $\pm$ 0.0165 & 104.00 $\pm$ 0.05 & 0.82 $\pm$ 0.03 \\
+10 bps & \ourm{} & 0.0505 $\pm$ 0.0016 & \textbf{130.12 $\pm$ 1.40} & \textbf{1.02 $\pm$ 0.00} \\
\bottomrule
\end{tabular}}}
\label{tab:cost_sensitivity}
\end{table}

As an interpretability diagnostic, we track the policy-library evolution on NAS100 for seed 0.
After pretraining, the initial library contains 12 policy vectors.
During the subsequent 37 detected regimes, the framework performs 10 insertions, 12 merges, and 15 discards, leading to a final library of 22 vectors.
The final gate weights are also highly concentrated: the top five vectors account for 34.3\%, 26.1\%, 10.5\%, 10.2\%, and 4.3\% of the total weight, respectively.
This indicates that the library remains compact while preserving a small set of dominant reusable policies.

\section{Limitations and Future Work}\label{app: limitations_future_work}

While \ourm{} demonstrates strong performance and adaptability in dynamic financial markets, several limitations remain.
First, the regime detection module relies on statistical change-point detection, which may struggle with subtle transitions or gradual market changes, especially in high-frequency settings where noise obscures regime boundaries.
Second, the current policy-library maintenance strategy is based on cosine similarity, which is efficient but may merge distinct policies with similar parameters or discard rare yet useful regimes. Third, although the regime-gate module provides interpretability through attention weights, stronger explainability and financial transparency are still needed for real-world deployment. Fourth, our experiments focus on day trading, so the generalization of \ourm{} to higher-frequency data remains to be validated. Finally, although \ourm{} performs strongly over long investment horizons, it may be less effective in short-horizon settings that require very rapid adaptation.

Future work includes more data-driven regime segmentation, broader cross-market evaluation, improved interpretability and risk control, and extensions that better support short-horizon trading.

\end{document}